\documentstyle[]{mn}
\ifnfsstwo

\fi

\ifnfssone
  \newmathalphabet{\mathit}
    \addtoversion{normal}{\mathit}{cmr}{m}{it}
    \addtoversion{bold}{\mathit}{cmr}{bx}{it}

\fi

\ifoldfss

\fi

\loadboldmathitalic
\loadboldgreek
                                                              
\def\msun{\thinspace\hbox{$\hbox{M}_{\odot}$}}

\def\lsun{\thinspace\hbox{$\hbox{L}_{\odot}$}}
\def\gal{galaxy}
\def\gals{galaxies}
\def\el{elliptical}
\def\ab{abundance}
\def\ev{evolution}
\def\chev{chemical evolution}
\def\for{formation}
\def\sfor{star formation}

\def\cf{cooling flow}
\def\gw{galactic wind}
\def\lum{luminosity}

\title[Elliptical Galaxies and QSOs]
       {Formation and Evolution of Elliptical Galaxies and QSO Activity}
\author[A. C. S. Fria\c ca and R. J. Terlevich]
       {Amancio C. S. Fria\c ca$^1$ and Roberto J. Terlevich$^2$\\
        $^1$Instituto Astron\^omico e Geof\'\i sico, USP,
        Av. Miguel Stefano 4200, 04301-904 S\~ao Paulo, SP, Brazil\\
        $^2$Royal Greenwich Observatory,
        Madingley Road, Cambridge CB3 0EZ, UK}
\pubyear{1996}

\begin{document}
\maketitle

\begin{abstract}
We present the results of a numerical code that combines 
multi-zone chemical evolution with 1-D hydrodynamics
to follow in detail the evolution and radial behaviour 
of gas and stars during the formation of elliptical galaxies.
We use the model to explore the links between the
\ev\ and formation of elliptical galaxies and QSO activity.
The  knowledge of the radial gas flows in the galaxy allows us to
trace metallicity gradients,
and, in particular, the formation of a high-metallicity core in ellipticals.
The high-metallicity core is formed soon enough to explain the metal
abundances inferred in high-redshift quasars.
The star formation rate and the subsequent feedback regulate the
episodes of wind, outflow, and cooling flow, thus affecting the recycling
of the gas and the chemical enrichment of the intergalactic medium.
The \ev\ of the galaxy shows several stages, some of which are characterized
by a complex flow pattern, with inflow in some regions and outflow in other regions.
All models, however, exhibit during their late \ev\ a \gw\ at the outer boundary
and, during their early \ev, an inflow towards the \gal\ nucleus.
The characteristics of the inner inflow could explain 
the bolometric luminosity of a quasar lodged at the galaxy
centre as well as the evolution of the optical luminosity of quasars.
\end{abstract}


\section{Introduction}

Imaging studies of the faint extensions around QSOs indicate that \el\ \gals\ are
the host \gals\ of the radio-loud and the brightest QSOs 
(Smith et al. 1986; Hutchings, Janson \& Neff 1989;
Hutchings et al. 1994; Hutchings \& Morris 1995; McLeod \& Rieke 1995;
Aretxaga, Boyle \& Terlevich 1995;
Disney et al. 1995; Bahcall, Kirkakos \& Schneider 1996;
Ronnback et al. 1996; Taylor et al. 1996).
As a matter of fact, at high redshifts ($z  > 2$),
the only galactic systems available to harbour QSOs are the spheroids,
since the disks are formed much later.
In addition, the epoch of completion of the large spheroids ($z  > 2$)
coincides with the peak in the QSO activity ($2 < z < 3$)
(Schmidt et al. 1991), suggesting a relation between the QSO phenomenon
and the \for\ of large \el s.
In fact, in recent years there has been increasing evidence linking
QSO activity with galaxy formation. The high metal content
of high redshift QSOs, the
high dust content (several $10^8$ \msun\ of dust) of 
distant QSOs (Andreani, Franca \& Cristiani 1993;
Isaak et al. 1994; McMahon et al. 1994; Omont et al. 1996)
plus the possible relation between the galaxy
luminosity function (LF) and the QSO LF 
(Terlevich \& Boyle 1993, hereafter TB93; 
Boyle \& Terlevich 1998, hereafter BT98)
provides tantalising evidence of this link. 
In any case, the fact that even the highest
redshift QSO has strong metal lines in its spectrum requires 
that the broad line region (BLR) gas has been enriched by a stellar population
formed before $z\sim 5$.
Work by TB93 and Hamann \& Ferland (1993) (hereafter HF93) 
highlighted the importance of metal production
in the early evolution of a galaxy. 
Since QSOs are seen up to redshifts of $z\sim 5$, the supersolar
metallicities required by the BLR models should be reached by $\sim 1$~Gyr
since the beginning of the galaxy formation epoch. 
This evolutionary time scale is
an important constraint in chemical enrichment models.

Two fundamental relations involving intrinsic parameters of
elliptical galaxies show remarkable little dispersion
and point towards an early formation of ellipticals.
They are the
colour-luminosity relation and the ``fundamental plane'' relating the
total luminosity of an elliptical to its central velocity dispersion
and surface brightness.  The tightness of the colour-luminosity relation
provides strong evidence that most of the present stellar population was
formed at $z > 2$ (Bower, Lucey and Ellis 1992). The narrowness of the
``fundamental plane'' gives additional support to that conclusion (Renzini
and Ciotti 1993) and in addition indicates that the properties of the
core (velocity dispersion, Mg$_2$ strength) are intimately linked 
to the galaxy global ones (D$_n$, luminosity).  
There is also good evidence that the stellar population in
massive ellipticals is metal rich with respect to the Sun, and shows large
radial metallicity gradients (e.g., Worthey, Faber and Gonzalez, 1992;
Davies, Sadler and Peletier 1993).  
These properties of the \el\ \gals\ may also be related to
the fact that they probably harbour QSOs in their centres
at some stage of their \ev.

One-zone chemical evolution models have been used (Hamann \& Ferland
1992, HF93, Padovani \& Matteucci 1993, Matteucci \& Padovani 1993)
to investigate the chemical history and the fueling of QSOs.
However, during the early evolution of the elliptical galaxy,
it is expected several episodes of gas outflow and inflow 
which cannot be followed by the one-zone model.
In addition, 
the one-zone chemical evolution models that attempt
to explain the high metal content in high redshift QSOs tend to
overproduce metals (averaged over the entire galaxy) and
predict an excessively high luminosity for the parent galaxy of the QSO.
For example, HF93 model M4 reproduces the rapid metal
production needed but it is overluminous. 
In this model, an elliptical of $10^{11}$ \msun\ has
a peak bolometric luminosity of $\sim 2 \times10^{13}$ \lsun\ 
at an age of $\sim 0.1$ Gyr. 
But  an elliptical with $10^{11}$ \msun\ is at present only a sub-$L^*$ galaxy
(with a blue luminosity of $\sim 0.3 L^*$, for $M_B^*=-21$
and $[M/L_B]=10$), so that
for the most luminous ($M_B \approx -24$) ellipticals in the nearby Universe
HF93 model M4 predicts luminosities of up to
$10^{15}$ \lsun\ during its formation.
These luminosities are higher than the QSO luminosities!!

Note that only the core of the galaxy has to be metal rich in accordance
with the observed metallicity gradients in nearby galaxies.
In the starburst model for QSOs (TB93 and references therein), 
the QSOs are the young cores of massive ellipticals
forming most of the dominant metal-rich population in a short starburst.
The core mass, which participates in the starburst,
comprises only a small fraction ($\sim 5$ \%)  of the total galaxy mass.
Also in the standard supermassive black hole model for QSOs, only $0.5-1$ per cent
of the \gal\ mass goes into the black hole (Haehnelt \& Rees 1993).
The excessive production of energy and metals in the one-zone model
arises, therefore, from its inability to resolve the core of the \gal.

The QSO LF undergoes strong \ev\ between $z=2$ and the present epoch, 
with the redshift
dependence of the LF being well-described by a constant comoving space
density  and a pure power-law luminosity \ev\ $L(z) \propto (1+z)^k$
(Boyle et al. 1988, 1991, BT98), with $k$ in the range $3.1<k<3.6$.
This luminosity \ev\ is linked to the mass flow into the galactic nucleus
hosting the QSO, both in the supermassive black hole scenario and
in the starburst model for AGN.
Since the luminosity is expected to be proportional to the mass accretion rate
into the nucleus,
in order to make predictions about the luminosity \ev\ of the QSO
one needs the \ev\ of the central gas inflow.
Again, this piece of information is not provided by the one-zone model.

In view of the limitations of one-zone chemical models, we have developed 
a multi-zone chemo-dynamical model
that self-consistently combines chemical evolution and
numerical hydrodynamics.
In this paper, we use this code to investigate a number of topics related to
\for\ and \ev\ of \el\ \gals\ and the link young \el s-QSOs:
1) the formation of a high-metallicity core in ellipticals;
2) the radial metallicity gradients in ellipticals;
3) the chemical enrichment of the intracluster medium by ellipticals;
4) the evolution of the luminosity of young cores of \el s/QSOs.
The approach of the present work is first to develop a realistic sequence
of models which reproduces the main properties of \el\ \gals.
We then investigate the \ev\ of the \gal\ core and of the gas inflow into
the nucleus and, within the scenario in which the galaxy nucleus
hosts a QSO, we compare the predictions of the model with QSO
observations.

The chemo-dynamical evolution code is described in Section 2.
Section 3 presents the evolution of the interstellar medium (ISM) and of the \sfor.
Section 4 considers the chemical enrichment of the intracluster
medium by \gw s from elliptical galaxies.
Section 5 is devoted to \ab s  and metallicity gradients 
of the stellar population in ellipticals. 
The predictions of our models for the \ev\ of the QSO LF are presented in Section 6.
Some  concluding remarks are given in Section 7.

\section{The chemo-dynamical model}

Apart from spectral-synthesis models not discussed in this paper,
two classic approaches have been used to study the \for\ and \ev\ of elliptical
galaxies: hydrodynamical models aimed at reproducing the evolution of
the X-ray emitting gas in elliptical galaxies
(David et al. 1990,1991; Ciotti et al. 1991);
and one-zone chemical evolution models, which seek to reproduce
the global evolution of gas and stars, giving results on the evolution
of metal abundances (Arimoto \& Yoshii 1987, Matteucci \& Tornamb\'e
1987, Padovani \& Matteucci 1993, Matteucci \& Padovani 1993).

Unfortunately, both models give only a restricted description of the
first few Gyrs of the elliptical galaxy, the stage we regard as more interesting
in connection to the elliptical \gals-QSOs link.
In the case of the above hydrodynamical models, 
one central problem is that the gas or ISM represents
a major component of the young galaxy, which not only reacts
to the input of energy of stars but also, via \sfor, builds up the stellar population.
Moreover, the deposition of energy, mass and metals into the ISM
results from former star generations, 
and thus in a complete computation of the stellar input into the gas 
the whole star formation history should be stored.

Regarding the one-zone chemical evolution models, they do not
follow the dynamical evolution of the gas, and, in particular,
the evolution of the galactic wind, a central ingredient in
many of these models. 
The gas distribution and velocity field are not described by these
models, since the situation is far more complex than the existence or
not of a galactic wind. The gas flow is not necessarily coordinated
along the galaxy; there could be
inflow at one radius and outflow at another radius.
In addition, these models do not give information on
the space distribution of the stars formed  and
metallicity gradients. This latter issue is
needed for the understanding of the evolution of the core of the ellipticals, 
the region which is most readily observable,
where the metal-rich stars are located, and which lodges the QSO.

A more promising approach is represented by the chemo-dynamical model,
which combines hydrodynamical and chemical evolution.
Such models have been employed to study \el\ \gals\ (Burkert \& Hensler 1989;
Theis, Burkert \& Hensler 1992) and spiral \gals\ (Burkert, Truran \& Hensler 1992;
Samland, Hensler \& Theis 1997).
The chemo-dynamical model avoids a number of uncertainties of the one-zone models.
One common assumption, for example, is that star formation ceases after
the system has evolved long enough.  For instance, in HF93 models,
star formation stops when the gas mass fraction drops below 3\%.  This in
principle recognizes the existence of a density threshold for star
formation (Kennicutt 1989), but the choice of the value of the critical
density (or gas fraction) is arbitrary.  On the other hand, Matteucci
\& Padovani (1993) models assume, as most of the chemical evolution
models do, that the star formation stops when the galactic wind starts.
The existence of galactic winds in ellipticals was first suggested by
Mathews \& Baker (1971) in order to explain the apparent lack of gas in
these systems.  Later, Larson (1974b) invoked galactic winds to reproduce
the relation mass-metallicity in ellipticals.  Following Larson,
the condition for the onset of a galactic wind in Matteucci \& Padovani
(1993) models is that the thermal energy content of the gas heated by
SNe exceeds the binding energy of the gas.
Note, however, that the suppression of star formation by galactic winds
assumes that the wind instantly devoids the galaxy of gas.
A more realistic approach should include the time delay
due to the finite flow velocity of the gas leaving the galaxy.
Moreover,
winds will not necessarily curtail all star formation, as can be seen from
observations of starburst galaxies which can have simultaneously kpc-scale
X-ray wind and ongoing massive \sfor\ (Heckman, Armus \& Miley 1987).

In the present work we use the chemo-dynamical model
developed by Fria\c ca \& Terlevich (1994) to follow 
the \ev\ of \el\ \gals\ from the protogalaxy stage.
The model combines
a multi-zone chemical evolution solver and a 1-D hydrodynamical code.
The elliptical galaxy, assumed to be spherical, is
subdivided into several spherical zones and
the hydrodynamical evolution of its ISM is calculated 
(see Fria\c ca 1986, 1990, 1993). 
Then, taking into account the gas flow, 
the chemical evolution equations are solved for each zone.
In this work, 100 zones are used, with the inner
boundary at 100 pc and the outer boundary at the galaxy tidal radius.
A total of $\approx 100$ star generations over 13 Gyr is stored for the chemical evolution
calculations.
The length of the star generations increases with the stellar age,
ranging from $10^6$ yr for newly formed stars to
$\approx 3\times 10^8$ yr for the oldest stars.
Our models include inhibition of star formation when the density
is too low or when the gas is expanding, 
but the star formation is never sharply cut off
and continues after the first galactic wind has been established.
This allows us to study the whole star formation history of the galaxy,
as well as the development of both late wind
and late cooling flow episodes.

\subsection{The hydrodynamical evolution}

The hydrodynamical evolution of the spherically symmetric ISM is given by
solving the fluid equations of mass, momentum and energy conservation
\begin{equation}
{\partial\rho\over\partial t}+{1\over r^2}{\partial\over\partial r}(r^2 \rho u)
= \alpha \rho_* - \nu \rho ,
\end{equation}
\begin{equation}
{\partial u\over\partial t}+u{\partial u\over\partial r}
=-{1 \over \rho}{\partial p\over\partial r}
-{GM \over r^2}-\alpha{\rho_* \over \rho}u,
\end{equation}
\begin{eqnarray}
{\partial U \over \partial t}+u{\partial U \over \partial r}
&=&{p \over \rho^2}
\left({\partial \rho \over \partial t}+u{\partial \rho \over \partial r}\right)
-\Lambda\rho \nonumber \\ 
 & &\mbox{}+\alpha{\rho_* \over \rho}\left(U_{inj}+{u^2 \over 2}-U-{p \over \rho}\right),
\end{eqnarray}
where $\rho$, $u$, $p$, and $U$ are the gas density, velocity, pressure and
specific internal energy.
The equation of state $U=(3/2)p/\rho$ completes the system of equations.
The total binding mass $M$ is the sum of three components: gas, stars and a dark
halo, $M_g$, $M_*$, $M_h$, respectively. The gas and the stars exchange mass 
through star formation and stellar mass losses (supernovae, planetary
nebulae, and stellar winds). The dark halo has no interplay with the gas and
the stars, and it is given by a static mass density distribution
$\rho_h(r)=\rho_{h0}[1+ (r/r_h)^2]^{-1}$,
where $\rho_{h0}$ is the halo central density and $r_h$ is the halo core radius.

The star formation process and the restoring of mass to the ISM by dying stars
are represented by the specific rates for star formation $\nu$
and for gas return $\alpha$.
The terms $\alpha \rho_*$ and $\nu \rho$ in the continuity equation
couple the gas density to the stellar mass density $\rho_*$.
It is expected that the stars do not remain in the zones in which they are formed.
In addition, the final positions to where they have moved
should reproduce the density profile of a realistic galaxy.
We take this into account by 
radially moving each zone of newly formed stars
in one free-fall time to a final position
such as the resulting stellar mass distribution follows a King profile
$\rho_*(r)=\rho_{*0}[1+(r/r_c)^2]^{-3/2}$,
where $\rho_{*0}$ and $r_*$ are
the central stellar density and the stellar core radius, respectively.
Both the stellar distribution and the dark halo are truncated 
at a common tidal radius $r_t$,
beyond which there is no \sfor.

The stars are assumed to die either as supernovae (SNe) or as planetary nebulae,
when instantaneous ejection of mass and energy occurs.
The energy per unit mass injected into the gas by the dying stars 
can be divided in three contributions due to Type I SNe, Type II SNe and
quiescent stellar mass loss (planetary nebulae and stellar winds)
\begin{eqnarray}
U_{inj} &=&(\alpha_{SNI} E_{SNI}/M_{SNI}+\alpha_{SNII} E_{SNII}/M_{SNII}
 \nonumber \\
 & &\mbox{}+\alpha_* U_{inj,*})/\alpha,
\end{eqnarray}
where $\alpha_{SNI}$, $\alpha_{SNII}$, and $\alpha_*$ are the specific
gas return rate by Type I SNe, Type II SNe, and quiescent stellar
mass loss, respectively ($\alpha=\alpha_{SNI}+\alpha_{SNII}+\alpha_*$).
$M_{SNI}$ and $M_{SNII}$, $E_{SNI}$ and $E_{SNII}$ are the mass and kinetic
energy of the supernova ejecta, respectively. Here, Type I SN stands for
Type Ia only, whereas Type Ib is included in the Type II SN.
The gas which is lost from stars as wind or planetary nebulae is assumed to be
thermalized to the temperature given by the velocity dispersion of the stars
(i.e., $U_{inj,*}=(3/2)\sigma_*^2$, where $\sigma_*$ is the
one-dimensional stellar velocity dispersion).

The adopted cooling function $\Lambda(T)$, defined so that $\Lambda(T)\rho^2$
is the cooling rate per unit volume, takes into account the variations of
the gas abundances predicted by the chemical evolution calculation.
For the sake of simplicity, instead of considering the abundances of all
elements included in the chemical evolution calculations,
our cooling function depends only the abundances
of O and Fe, which are the main coolants for $T>10^5$ K.
In the evaluation of the cooling function, the abundances of elements other
than Fe and O have been scaled to the O abundance as
$y_i=y_{i,P}+(y_{i,\odot}-y_{i,P}) y_O/y_{O,\odot}$,
where $y_i$ is the abundance by number of the element $i$ 
(the cooling function takes into account emission from H, He, C, N, O, Ne, Mg,
Si, S, Cl, Ar and Fe), and $y_{i,P}$ and $y_{i,\odot}$
are the primordial (i.e., Y=0.24 and Z=0.0) and solar  (Grevesse \& Anders 1989)
abundances of the element $i$.
The atomic database used in the determination of the cooling function comes
from the photoionization code AANGABA (Gruenwald \& Viegas 1992).
A novelty of these models is the  self-consistency of
the hydrodynamics, chemical evolution and atomic physics,
since the cooling function is evaluated based on the actual
chemical abundances obtained from the chemo-dynamical modelling.

The spherically symmetric hydrodynamics equations are solved using a 
finite-difference, implicit code based on Cloutman (1980).
The code is run in the Euleurian mode and the grid points are
spaced logarithmically. The grid has between 150 and 300 cells,
the first cell being 50 pc wide.
The innermost cell edge is located at 100 pc and the outer boundary
at twice the tidal radius of the galaxy.
The artificial viscosity for the treatment of the shocks follows the
formulation of Tsharnutter \& Winkler (1979)
based on the Navier-Stokes equation. 
In contrast with the von Neumann--Richtmyer artificial viscosity
(Richtmyer \& Morton 1967), this form of artificial
viscosity vanishes for an homologous contraction. 

The outer boundary conditions on pressure and density are derived by
including an outer fictitious cell, the density and pressure in which
are obtained from extrapolation of power laws over the radius
fitted to the five outermost real cells.
The inner boundary conditions adjust according to whether  inflow
or outflow prevails locally. 
During inflow, the velocity at the inner boundary is extrapolated from
the velocities at the innermost cell edges
whereas during outflow, the velocity at the inner boundary is set to zero.

The initial conditions assume an entirely  gaseous
($M_*=0$) proto\gal.
The gas has temperature $T_0=10^4$ K, primordial chemical abundances
and density distribution following that of the dark halo.
The models are evolved until the present epoch ($t_G=13$~Gyr). 
A $H_0=50$~km~s$^{-1}$~Mpc$^{-1}$, $\Omega=1$ cosmology is adopted
throughout this paper.

\subsection{The chemical evolution}

Chemical \ev\ occurs as stars form out of the ISM
evolve and eject gas back into the ISM
via stellar winds, planetary nebulae and SNe. 
In this work, we use conventional chemical \ev\ formulations 
frequently used in classic one-zone models
in order to make it easier comparison with previous works.
Within each spherical zone, in which the model galaxy has been divided,
the evolution of the abundances of six chemical species (He, C, N, O, Mg, Fe) 
is calculated by solving the basic equations of chemical evolution
(Talbot \& Arnett 1973, Tinsley 1980, Matteucci \& Tornamb\`e 1987).
We do not assume instantaneous recycling approximation for the chemical
enrichment, but instead the delays for gas restoring from the stars
are taken into account by using main-sequence lifetimes $t_m$ (in Gyr)
$\log m=0.0558\log^2 t_m -1.338\log t_m +7.764$ 
for stellar masses $m \leq 6.6 \msun$ (Renzini \& Buzzoni 1986),
and $t_m = 1.2 m^{-1.85} +0.003$ for $m> 6.6 \msun$
(G\"usten \& Mezger 1983).
Instantaneous mixing with the ISM is assumed for the stellar ejecta. 

We express the relation of the \sfor\ rate (SFR) to the gas content,
by writing the SFR  as $\psi(r,t)= \nu \rho$,
where $\nu$ is the specific SFR. 
Unfortunately, there is no widely accepted theory of star formation.
Thus, following Schmidt (1959, 1963), one usually adopts 
a dependency of $\nu$ on a power of the gas density,
\begin{equation}
\nu= \nu_0 (\rho / \rho_0)^{n_{SF}}.
\end{equation}
$n_{SF}=0$ and $n_{SF}=1$ for the linear ($\psi\propto\rho$) and 
the quadratic ($\psi\propto\rho^2$) models of Schmidt.
It is not expected  that the SFR  strictly follows a power law of the density;
the parameter $n_{SF}$ is meant only to describe the nonlinearity  
of the dependence of the SFR on the gas content.
There are, however, good theoretical and observational reasons for believing that
the SFR depends at least linearly on the gas content.
Kennicutt (1989) finds that the SFR per unit area of disk galaxies
varies approximately as the 1.3 power of the gas surface density.
Chemical evolution models for the Galaxy require a weak dependence
of the SFR on the gas density: 
a range $n_{SF}=0-0.5$ is found by Rana \& Wilkinson (1986), whereas
Matteucci \& Fran\c cois (1989) favor $n_{SF}=0.1$ over $n_{SF}=1$.
Some values of $n_{SF}$ suggested for idealized cases bracket the observational
estimates above.
In this paper, most of the models have 
the so-called {\it standard SFR}, i.e., $n_{SF}=1/2$,
in which the time scale for \sfor\ is proportional to the local dynamical time
(Larson 1974a).
Models have also been run for 
a $n_{SF}=1/3$ SFR, 
which corresponds to a cloud collision model for star formation
with constant filling factor of the star-forming clouds
(Negroponte \& White 1983),
and a linear SFR ($n_{SF}=0$) (the simplest case).
The normalization  in eq. (5)
was set as $\nu_0=10$~Gyr$^{-1}$ 
in agreement with the $\sim 10^8$ yr time scale for \sfor\
required by chemical evolution models
in order to reproduce the supersolar [Mg/Fe] ratio in giant \el s
(Matteucci \& Tornamb\'e 1987; Matteucci 1992).
The fiducial value for the gas density $\rho_0$ is taken as the initial average gas
density inside the halo core radius.
We included inhibition of star formation for expanding gas
($\nabla.u>0$) or inefficient cooling
(i.e., for a cooling time $t_{coo}=(3/2)k_B T/\mu m_H \Lambda(T) \rho$
is longer than the dynamical time $t_{dyn}=(3\pi/16\,G\,\rho)^{1/2}$)
by multiplying $\nu$ as defined in eq. (5) by the inhibition factors
$(1+t_{dyn}\,{\rm max}(0,\nabla.u))^{-1}$ and $(1+t_{coo}/t_{dyn})^{-1}$.

The stellar birthrate (per unit volume) of stars of mass $m$
is given by $\psi(r,t)\phi(m)$, where $\phi(m)$ is
the initial mass function by number (IMF).
The IMF is assumed to be independent
on time and position and given by $\phi(m)=C\,m^{(1+x)}$. 
Here we adopt a Salpeter IMF ($x=1.35$), 
normalized over the mass range $m=0.1-100$~$\msun$.

The gas restoring from the stars 
depends on the initial stellar mass. Single stars in the mass range
$0.1<m/\msun<8$ end their life as planetary nebulae and leave  
helium or C-O white dwarfs with masses smaller than 1.4 $\msun$. 
Single stars with masses above 8 $\msun$ end their life as Type II SNe. 
No distinction is made between Type II and
Type Ib, Type Ib being considered a Type II which has lost its envelope
prior to the explosion. Type Ia supernovae are assumed to originate from
binary systems of total mass in the range $3<m/\msun<16$ in which the
primary evolves until it becomes a C-O white dwarf. Mass transfer from the
slower-evolving secondary triggers C-deflagration onto the primary
when the latter reaches the  Chandrasekhar mass (Whelan \& Iben 1973).
The computation of the Type Ia SN rate follows Greggio \&
Renzini (1983). An important parameter in this scenario is $A_{SN\,I}$, the mass
fraction of the IMF between $3<m/\msun<16$ that goes to binary systems
giving rise to Type Ia SNe. Following
Matteucci \& Tornamb\'e (1987) and Matteucci (1992), we have chosen 
$A_{SN\,I}=0.1$. 
This value should be verified {\it a posteriori} by checking the predicted
against the observed SN Ia rate in ellipticals.

The nucleosynthesis prescriptions for
the envelopes of single intermediate mass stars
($0.8<m/\msun<8$) are taken from Renzini \& Voli (1981) 
(their case with $\alpha_c=0$, $\eta=0.33$).
We consider production of secondary N via CNO shell burning in the
envelopes of massive stars ($m>8$ $\msun$) following the case B
of Talbot \& Arnett (1973) (100\% conversion of the envelope C and O into N).
For stars in the mass range $8-10\msun$, the nucleosynthesis prescriptions
are those suggested by Hillebrandt (1982,1985).
The yields from Type II SNe with progenitors with mass between 10 and 40
$\msun$ are from Weaver \& Woosley (1993) and Woosley \& Weaver (1995),
using a value of the $^{12}C(\alpha,\gamma)^{16}O$ reaction rate
which is 1.7 times the Caughlan \& Fowler (1988) value.
The results from the nucleosynthesis calculations for stars more massive
than $\approx 18$ $\msun$
depend critically on the chosen value of this reaction rate.
Since Woosley \& Weaver (1995) do not calculate models 
for SN II progenitors more massive than 40 $\msun$, 
we extrapolate their yields for the mass range 40-100 $\msun$ 
based on the trend of yields with increasing mass
derived from Woosley \& Weaver (1986), who use a higher value for the
$^{12}C(\alpha,\gamma)^{16}O$ reaction rate.
For masses above 30 $\msun$, the results also depend on the assumed
kinetic energy at infinite $KE_{\infty}$ of the piston of the explosion.
If $KE_{\infty}$ is not large enough, there is some fall-back of material,
and a more massive remnant (a black hole) is formed,
locking the nucleosynthesis products and reducing the yields, specially
that of the iron.
Larger values of $KE_{\infty}$ prevent the fall-back and lead to a larger
amount of ejected iron.
The sequence A of models of Woosley \& Weaver has 
nearly constant $KE_{\infty}$, so that fall-back is not prevented,
a black hole is formed and little iron is ejected.
In this work, we use the sequence with $KE_{\infty}$ increasing
with mass (models S30B, S35C and S40C),
resulting in larger Fe yields and smaller remnants.
Finally,
the yields from Type Ia SNe are from Nomoto et al.  (1984) (their model W7).

We adopt the initial mass--final mass relation given by Iben \& Renzini (1983)
for the white dwarf remnants of 
single stars with initial mass in the range $0.1< m/\msun<8$.
Stars more massive than 8 \msun, which explode as Type II (or Ib)
supernovae are assumed to leave 1.4 \msun\ neutron stars remnants.
The Type Ia supernovae are assumed to leave no remnant.

 \begin{figure}
 \vspace{86mm}
 \caption{The \ev\ of abundance ratios for our reference one-zone model
illustrating the effect of the convection parameter $\alpha_c$ of the models
of Renzini \& Voli (1981) for the yieds of intermediate mass stars.
Throughout this paper, the bracketed ratios denote the ratio
of mass abundances relative to their solar values.}
 \end{figure}

We performed several checks of the chemical \ev\ solver of our code
by comparing the results of one-zone runs with classic \gw\ one-zone models
for \el\ \gals\ (Arimoto \& Yoshii 1987, Matteucci \& Tornamb\'e 1987,
Matteucci 1992).
Here we consider a reference one-zone model,
similar to the model A2 of Matteucci (1992),
except for some differences in
the nucleosynthesis prescriptions and stellar lifetimes.
Our model represents a \gal\ initially gaseous
with mass $M_G=10^{12}$ \msun, and no dark halo. 
It is assumed a linear \sfor\ law, i.e. $\nu_{SF}=\nu_0=$ constant.
Following the relation $\nu=8.6 (M_G/10^{12}\,\msun)^{-0.115}$ Gyr$^{-1}$
derived by Arimoto \& Yoshii (1987),
$\nu_0$ is set as 9 Gyr$^{-1}$.
The model belongs to the class of SN-driven \gw\ models, in which
a \gw\ is established at a time $t_{gw}$.
At this time, the gas is swept from the \gal\ and 
the \sfor\ is turned off forever.
The condition for the onset of the \gw\  is that
the thermal content of the SN remnants of the \gal\  equals
the gravitational binding energy of the \gal\ ISM
(the computation of these two quantities follows Matteucci 1992).
Our model gives $t_{gw}=0.81$ Gyr, in good agreement with
$t_{gw}=1.01$ Gyr of model A2 of Matteucci (1992).

 \begin{table*}
 \begin{minipage}{160mm}
  \caption{Galaxy mass and wind evolution}
  \begin{tabular}{@{}lccccccccc}
Model & $r_h$ & $M_*(t_G)$ & $M_g(t_G)$
& $t_w$ & $t_{w,e}$ & $M_*(t_w)$ & $M_g(t_w)$
& $M_w$ & $M_{w,Fe}$ \\
 & (kpc) & ($10^{11}\msun$) & ($\msun$) 
& (Gyr) & (Gyr) & ($10^{11}\msun$) & ($10^{11}\msun$) 
& ($10^{11}\msun$) & ($10^{8}\msun$) \\[3pt]
1 & 2.5 & 1.32 & 1.70$\times 10^7$ & 1.09 & 1.49 & 1.39 & 0.58 & 0.72 & 2.74 \\
2 & 3.5 &  2.38 & 4.73$\times 10^7$ & 1.17 & 1.60 & 2.53 & 1.28 & 1.42 & 4.93 \\
5 & 5 &   5.90 & 2.76$\times 10^8$ & 1.25 & 1.99 & 6.05 & 3.36 & 4.22 & 12.7 \\
10 & 7 & 11.6 & $2.06\times 10^9$ & 1.51 & 2.68 & 11.5 & 6.73 & 6.45 & 25.1 \\
20 & 10 & 23.7 & $2.77\times 10^{10}$ & 1.65 & 3.56 & 22.0 & 14.0 & 12.0 & 31.6 \\
50 & 15 & 66.4 & $2.14\times 10^{11}$ & 2.22 & 5.40 & 59.3 & 34.0 & 18.2 & 49.8 \\
2(1/3) & 3.5 &  2.83 & $5.23\times 10^7$ & 1.32 & 1.69 & 3.03 & 0.94 & 1.10 & 5.60 \\
2(0) & 3.5 &  1.33 & $1.49\times 10^7$ & 0.044 & 0.373 & 0.65 & 1.34 & 0.68 & 3.96 \\
  \end{tabular}
\end{minipage}
\end{table*}

Fig. 1 shows the \ev\ of abundances
of C, N, O, Mg and Fe for our reference model.
Our results for the abundance \ev\ are similar to those of
the model M4 of HF93
and the model of Matteucci \& Padovani (1993) 
with luminous mass of $10^{12}$ \msun\ and Salpeter IMF
(note that HF93 assumes $\nu_0=6.7$ Gyr$^{-1}$ and $x=1.1$ for the IMF,
and Matteucci \& Padovani model includes a dark halo).
In order to illustrate the effect of the convection parameter 
$\alpha_c$ of the models of Renzini \& Voli (1981)
for the yields of intermediate mass stars,
we show runs for their cases
($\alpha_c=0$, $\eta=0.33$) and ($\alpha_c=1.5$, $\eta=0.33$).
The effect of $\alpha_c$ can be seen in the \ev\ of [N/H]
(throughout this paper, the bracketed ratios denote the ratio
of mass abundances relative to their solar values,
i.e., [X/H]$\equiv$(X/H)/(X/H)$_{\odot}$).
For $\alpha_c=0$, there is secondary only N production,
and for $\alpha_c=1.5$ there is some primary production of N
due to hot-bottom burning in the stellar envelope.
Earlier than $3\times 10^7$ yr, 
there is only secondary production of N by high mass stars
and the pace of N enrichment is roughly proportional to the gas metallicity.
For later times, the yield of stars with $m<8$ \msun\ is important,
and in the case $\alpha_c=0$, 
also the intermediate mass stars produces only secondary N,
so [N/H] has a smooth \ev\ all the way until several times $10^8$ yr,
when it reaches highly supersolar values.
However, in the case $\alpha_c=1.5$, 
[N/H] shows a steep rise at $\sim 3\times 10^7$ yr
due to the first primary N ejected by intermediate mass stars.
Also note the fast increase of the abundances of $\alpha$-elements (O, Mg),
ejected by Type II SNe, which reach solar abundances at $\sim 10^8$ yr.
By contrast, the enrichment of Fe is delayed with respect to that of O
([Fe/H] becomes solar at $\sim 3\times 10^8$ yr)
since Fe is mostly produced in Type Ia SNe, with long-lived progenitors.

\section{Evolution of the ISM and Star Formation Rate}

We have built a sequence of galaxy models parameterized
according to the total (initial) luminous mass inside the tidal radius,
$M_G=M_g+M_*$
(the luminous mass of the \gal\ is comprised by the stellar and the gas components,
which are readly observable, mainly in optical and X-ray bands, respectively).
The models are further characterized by $r_h$, $r_t$,
and the ratio of the halo to the (initial) luminous mass, $M_h/M_G$.
We have investigated a grid of runs with $M_G$ between $10^{11}$ 
and $5\times10^{12}$ $\msun$,
and $r_h$ in the range $2.5-15$ kpc.
We set $r_t=28r_h$ and $M_h/M_G=3$. 
The choice of the $r_h-r_t-M_G$ relation
is based on the scaling laws of Sarazin \& White (1987).
$\rho_{*0}$ and $r_c$ of the stellar distribution
are related to the central stellar velocity dispersion $\sigma_*$
by the virial condition $4\pi G \rho_{*0} r_c=9\sigma_*^2$.
In addition, the model \gals\ follow a Faber-Jackson relation,
$\sigma_*=200(L_B/L_B^*)^{1/4}$ km~s$^{-1}$ (Terlevich 1992),
with $L_B$ related to $M_G$ through $[M/L_B]=10$,
the mass-light ratio typical of an $L^*$ galaxy.
In this work, we have chosen
as fiducial model that one with $M_G=2\times 10^{11}$ $\msun$
and standard SFR.
Due to inflow and \gw\ episodes occuring during the \gal\ \ev,
its present  stellar mass is $\sim20$\% higher 
than the initial  $M_G$.
The fiducial model has 
$L_B=2.4\times 10^{10}$ $\lsun$, i.e., somewhat fainter than the break
luminosity of the Schechter luminosity function ($L_B^*=3.7\times 10^{10}$
$\lsun$), and, therefore, is representative of the population of elliptical \gals.
In this work, since we are particularly interested in the relation between
young \gals\ and QSOs, we will explore a mass range corresponding
to luminosities around and above $L^*$.

Table~1 summarizes the evolution of  the gas and stellar components of the galaxy.
The first column identifies the model: the number is $M_G$, the initial mass of the 
protogalaxy  in units of 10$^{11}$ $\msun$;
the number between parenthesis is $n_{SF}$ for models with $n_{SF} \neq 1/2$.
In this paper we will identify the models by the codes in the first column of Table 1
(e.g. model 2 is the fiducial model).
Column (2) gives $r_h$.
Columns (3) and (4) show
the present-day (at $t_G=13$ Gyr) stellar and gas masses of the \gal.
The following columns exhibit characteristics of the \gw\ which appears in all models:
$t_w$, the time of the onset of the \gw;
$t_{w,e}$, the end of the early wind phase, defined by the gas mass
being reduced to 10\% of its amount at $t_w$;
$M_*(t_w)$ and $M_g(t_w)$, the stellar and gas galaxy masses at $t_w$;
$M_w$ and $M_{w,Fe}$, 
the total and the iron masses ejected by the wind until $t_G$.

In this section, we investigate the \ev\ of the ISM
and the \sfor\ in our models.
We now proceed to follow in some detail the \ev\ of the fiducial model.
Since the initial conditions are out of equilibrium (the gas is initially cold), 
at the start of the calculations the gas falls towards the centre and is compressed,
giving rise to shocks that 
rapidly heat the gas in the core to approximately 
the virial temperature of the system ($T\sim 10^7$ K). 
Then, a highly efficient star formation is occurring throughout the galaxy
and a young stellar population is rapidly built up.

Following the initial violent star formation burst, the first Type II SNe appear
at $3.2\times 10^6$ yr,
and shortly after, the SN heating  dominates the energetics of the ISM.
Fig.~2 shows the evolution of the supernova rates over the whole \gal,
normalized to the initial protogalactic mass ($2\times 10^{11}$ $\msun$),
for the fiducial model.
The SN II rate reaches a maximum of
155 SNe (100~yr)$^{-1}$ $(10^{11}\msun)^{-1}$ at $3.9\times 10^8$ yr
and decreases rapidly after 1 Gyr.
The broad maximum of the SN II rate reflects the ongoing star formation
during the first Gyr of the galaxy.
The present SN II rate is negligible 
($3.8\times 10^{-4}$ SNe (100~yr)$^{-1}$ $(10^{11}\msun)^{-1}$ at 13 Gyr).
The Type~II SNe are followed by the Type~I SNe, which appear at $2.9\times 10^7$ yr
and attain a maximum rate of 19 SNe (100~yr)$^{-1}$ $(10^{11}\msun)^{-1}$
at $7.6\times 10^8$ yr. 
After 1 Gyr, the SN~I rate decreases less rapidly than the SN~II rate
and its value at $t_G=13$~Gyr is
0.20 SNe (100~yr)$^{-1}$ $(10^{11}\msun)^{-1}$. 
Given $M_*(t_G)=2.38\times10^{11}$ \msun\ and $[M/L_B]=10$,
this value corresponds to 0.17 SNU 
(1 SNU = 1 SN (100~yr)$^{-1}$ $(10^{10}\lsun_{B})^{-1}$)
and is bracketed by observational estimates of the SN~I rate:
0.22 $h_{50}^2$ SNU (Tammann 1982),
0.25 $h_{50}^2$ SNU (van den Bergh \& Tammann 1991),
0.03-0.09 $h_{50}^2$ SNU (Capellaro et al. 1993),
and 0.06-0.11 $h_{50}^2$ SNU (van den Bergh \& McClure 1994).
Only the SN I rate is significant in the later evolutionary
phases ($t>2$~Gyr) of the elliptical, when the star formation rate plummets
and the SN~II rate is drastically reduced.
The late evolution of the SN~I rate in the fiducial model 
can be approximated by a power law ($\propto t^{-1.77}$ from 1 to 13 Gyr).
However, the power law behaviour cannot be extrapolated to the first Gyr
of the model galaxy, since the plateau of the SN I rate between 0.5 and 1 Gyr
is not reproduced.

 \begin{figure}
 \vspace{86mm}
 \caption{The evolution of the supernova rates for the fiducial model.
The rates have been normalized to the initial protogalactic mass
($2\times 10^{11}$ $\msun$).}
 \end{figure}

The \ev\ of the gas velocity profile is shown in Fig. 3.
Three stages can be distinguished in the \ev\ of the \gal\ ISM.
The first stage is
a {\it global inflow} extending from the model inner boundary to the tidal radius.
At the beginning of this stage, the inflow rises from the settling of the gas
into the potential well of the dark halo.
When the central density has increased enough to allow radiative losses
to be important, a vigorous cooling flow is established in the inner kpc. 
Later during this stage, after the first SNe appear, 
the SN heating drives a wind at intermediate radii,
separating the global inflow into an inner cooling flow feeding the \gal\ nucleus
and an outer inflow falling towards the galaxy.
The wind advances towards the tidal radius and when it reaches the tidal radius at 
$t_w=1.17$ Gyr, 
the second stage begins, characterized by a {\it partial wind},
i.e., a cooling flow in the inner regions of the galaxy and a wind in the
outer regions. 
$t_w$ is to be identified with the onset of the \gw.
As more gas is consumed by star formation and expelled by the wind, the stagnation region
separating the inner cooling flow from the wind moves inwards, and 
at 1.8 Gyr the cooling flow has shrunk into the inner boundary,
so that a wind is established from 100 pc to the tidal radius,
characterizing the third, {\it total wind} stage.
The wind at the tidal radius reaches a peak velocity of 1860 km s$^{-1}$ at 2 Gyr, and
decreases afterwards, as the SN heating reduces due to the SN I rate decrease.

 \begin{figure}
 \vspace{86mm}
 \caption{The velocity profile of the gas for the fiducial model
at several evolutionary times.
The lines are labelled according to the time (in Gyr) they represent.
The top panel allows one to follow the global inflow stage
of the \ev\ of the ISM and
the lower panel shows the partial wind ($t=$ 1.2 and 1.6 Gyr)
and the total wind stages ($t=$ 2, 5, and 13 Gyr).}
 \end{figure}

Fig. 4 shows the \ev\ of the stellar mass and gas fraction for the fiducial \gal.
From the mass \ev\ results of Table~1 and Fig. 4, it is clear that the gas
is a major component of the \gal\ during  the first Gyr of \ev.
Gas is turned into stars by a high star formation rate,
but the gas consumption is not complete, since
the continuous gas infall constantly supply the \gal\ with gas.
For the fiducial model, the SFR is initially higher than the infall rate
leading to a decrease of the gas fraction 
as the stellar body of the \gal\ is built up.
The maximum SFR is $\sim 500$ $\msun$~yr$^{-1}$ at $6\times 10^8$ yr.
By then the stellar mass is $2\times 10^{11}$ $\msun$.
When the SFR has decreased to
$\sim 200$ $\msun$~yr$^{-1}$ at $8\times 10^8$ yr, 
the infall rate overtakes the SFR,
and the gas fraction reaches a minimum of 0.3 and then starts to increase again.
The increase of the \gal\ mass due to infall goes on until the onset of the \gw, at $t_w=1.17$ Gyr.
At this time, the \gal\ mass has almost doubled ($M_G(t_w)=3.8\times 10^{11}$ $\msun$),
and approximately one third of the \gal\ mass is in the form of gas.
The \gw\ that follows is so massive that almost all the remaining gas is removed
from the \gal. 
The gas expelled with the \gw\ represents a significant fraction
of the galaxy mass ($1.3\times 10^{11}$ $\msun$ from $t_w$ until 13 Gyr). 
The gas removal is very fast: 50\%(90\%) of the gas is expelled 
in $2.3\times 10^8$ yr ($4.3\times 10^8$ yr) after $t_w$.
In later stages of \ev, stellar losses continue to supply mass to the wind, 
which in turn depletes the gas.
During this epoch, the level of \sfor\ is very low, and, as a consequence,
newly formed stars do not replenish the dying stars, and the stellar mass decreases by 6 \%
from $t_w$ until now.
The net result of the early inflow, \gw\ and stellar losses is that the present-day
\gal\ mass is $\sim20$ \% higher than the initial protogalactic mass.
For other galaxy masses, there is also an increase of mass,
in the range of $15-30$\%,
but with no clear trend of $M_*(t_G)/M_G$ with $M_G$.
The model with $n_{SF}=1/3$ has a somewhat larger increase of 42\%.
The only discrepancy is represented by model 2(0), anomalous in many ways,
in which the \gw\ occurs very early and leads to a reduction in mass of 34\%.
In this model, due to the weak dependency of the SFR on the density,
the SFR is proceeding at high efficiency in the outskirts
of the \gal\, and the resulting SN heating avoids the massive initial infall
typical of the other models.

 \begin{figure}
 \vspace{86mm}
 \caption{Evolution of the stellar mass (solid line) 
and of the gas fraction (dotted line) for the fiducial model.}
 \end{figure}

Fig. 5 (top panel) shows for the fiducial model the \ev\ of the total SFR (over the whole \gal) 
and of the inner SFR (inside the inner kpc).
The total SFR initially has a value of 230 \msun\ yr$^{-1}$.
As more fresh gas of the early inflow falls into the \gal, 
the gas is compressed thus reducing the \sfor\ time scale.
Accordingly, the SFR rises up to 
a maximum of $480$ $\msun$~yr$^{-1}$ at $6\times 10^8$ yr.
During the first Gyr, the SFR is typically around 300 \msun\ yr$^{-1}$.
At $\sim 1$ Gyr, the depletion of gas by \sfor, 
reduces the SFR below 100 \msun\ yr$^{-1}$.
The onset of the \gw\ at 1.17 Gyr leads to
a drastic reduction of the SFR to $\sim 10^{-2}$ \msun\ yr$^{-1}$
in less than 0.5 Gyr.
While the total SFR shows more or less constant levels during the first Gyr of \ev,
the inner SFR (inside a radius of 1 kpc) shows dramatic variations, due the short
time scales for replenishment of gas and inhibition of \sfor.
It varies from 0.9 \msun\ yr$^{-1}$ at $t=0$ 
to a maximum of 70 \msun\ yr$^{-1}$ at $7\times 10^8$ yr,
with several brief starburst episodes.
Over the first Gyr of \ev, the inner SFR shows a typical range of
$20-50$ \msun\ yr$^{-1}$.
Note that, from the observational point of view,
the inner SFR is more interesting than the total SFR,
because only the inner regions of the \gal, due to their higher surface brightness,
would have \sfor\ activity detectable at high redshifts.

We have calculated for the fiducial model the \ev\ of the bolometric luminosity 
of the stellar population of the whole \gal\ (top panel of Fig. 5) 
by using the models of spectrophotometric \ev\ of Bruzual \& Charlot (1993).
Initially, the \lum\ increases steadly while a young stellar population
is being built up.
At $3.3\times 10^8$ yr, the young stellar population reaches  $10^{11}$ \msun. 
A maximum \lum\  of $2.3\times 10^{12}$ \lsun\ is reached at $6.1\times 10^8$ yr,
when the stellar mass is $2\times 10^{11}$ \msun.
The peak in \lum\ closely follows the maximum in SFR, since at this time the main
contribution to the \lum\ comes from high-mass stars.
It should be noted that
the luminosities of our models are more than one magnitude lower than
the luminosities in HF93. 
The HF93 model M4, intended to represent a giant \el, has a peak bolometric \lum\ of
$2\times 10^{13}$ \lsun\ at 0.1 Gyr for a \gal\ mass of $10^{11}$ \msun,
corresponding to a \lum\ of $4\times 10^{13}$ \lsun\ for the mass of the fiducial model.
Part of this discrepancy arises from the flatter IMF ($x=1.1$) assumed by HF93
in model M4, but most of it comes from the fact that HF93 models
consider that the whole \gal\ is involved in a starburst.
In model M4, the SFR reaches a pronounced maximum very rapidly at 0.1 Gyr
and then falls very rapidly.
The time scale for \sfor\ over the whole \gal\ is $\sim 10^8$ yr.
In our models, by contrast, the SFR keeps approximately the same levels
during the first Gyr of the galaxy.

 \begin{figure}
 \vspace{86mm}
 \caption{Top panel: \ev\ of the SFR over the whole \gal\ and
over the inner kpc (upper and lower solid lines, respectively), 
and the \ev\ of the bolometric luminosity of the stellar
population of the whole \gal\ in the fiducial model (dotted line).
Lower panel: \ev\ of the specific SFR averaged over the inner kpc (upper line)
and over the whole \gal\ (lower line).}
 \end{figure}

The lower panel of Fig. 5  shows 
the \ev\ of the average specific SFR inside the inner kpc and over the whole \gal.
The \sfor\ time scales in the inner region are much shorter than over
the whole \gal. 
At the beginning of the calculations the \sfor\ time scale is 1 Gyr over
the whole \gal\ and  $10^8$ yr in the inner kpc.
The average specific SFR over the whole \gal\ is typically between 1 and 3 Gyr$^{-1}$
during the first Gyr. 
Note that only the central region of the \gal\ is undergoing a violent starburst.
The average specific SFR inside 1 kpc is typically $\sim 30$ Gyr$^{-1}$
during the first Gyr.
The one-zone models of HF93 overestimate the \lum\ of the \gal, 
because they assume over the whole \gal\ 
a very short time scale of star formation ($\sim 10^8$ yr) in order to reproduce
the fast metal enrichment needed to account for the metal lines
observed in high redshift QSOs 
and the high [Mg/Fe] ratio in the nuclei of \el s.
But these constraints need to hold only in the core of the \gal.
In a multi-zone model, only the inner region of the \gal\ shows 
these very short time scales for star formation, 
while a relatively modest \sfor\ efficiency  characterizes the whole \gal, 
leading to much lower luminosities than in the one-zone models.

\section{The Galactic Wind}

In the early stages of the \ev\ of the \gal, the gas content is always high.
Gas is turned into stars by a high star formation rate,
but the gas consumption is never complete, because
the global infall constantly supply the \gal\ with gas.
The increase of mass due to infall goes on until the onset of the \gw.
The wind then removes almost all gas from the \gal.
As in the classic SN-driven wind models,
the onset of \gw\ occurs later for the more massive \gals
(see Table 1).
In addition, the wind removes gas slower in more massive \gals:
the duration of the early wind phase increases steadly from 0.40 to 5.4 Gyr,
as $M_G$ varies from $10^{11}$ \msun\ to $5\times 10^{12}$ \msun.
For all models, the wind occurs later than 1 Gyr.
The only exception is again model 2(0), for which 
the \gw\ occurs very early at $4.4\times 10^7$ yr.
For most of the models, the \gw\ persists until the present.
However, in the most massive models 20 and 50,
the very deep potential well causes the wind to be reversed 
to outer  inflow  at 11.8 and 8.8 Gyr, respectively,
so these models do not
exhibit \gw s at the present epoch.
Note that in a cluster environment, the higher pressures and densities
at the outer boundary could cause outer inflows even in
less massive \gals.

As shown in Fig. 6, the \gw\ can be very metal-rich. 
The central gas abundances soon become supersolar:
at $r=100$ pc, [O/H]$>1$ at $\sim 10^8$ yr,
and [Fe/H]$>1$ at $\sim 3\times 10^8$ yr.
After the wind has developed in the intermediate region of the \gal,
a metal-rich gas front advances outwards.
At the onset of the \gw\ ($t_w=1.17$ Gyr),
the gas metallicity at $r_t$ is low
([O/H]=$1.2\times 10^{-2}$ and [Fe/H]=$1.1\times 10^{-2}$),
although in the inner regions the gas is metal-rich
([O/H]=2.9 and [Fe/H]=5.6 at 1 kpc).
There is a delay between mass removal and metal removal by the wind.
When 90\% of the gas has been removed (at 1.59 Gyr), the metallicity  at
the tidal radius ([O/H]=0.8 and [Fe/H]=1.4) is still low in comparison to
the inner abundances ([O/H]=3.2 and [Fe/H]=8.9 at 1 kpc).
Only at 1.77 Gyr, the metal-rich gas front reaches the tidal radius
([O/H]=1.3 and [Fe/H]=8.1 at $r_t$), and a shallow
gas abundance gradient is established.
After this time, the \gw\ is very metal-rich.
The maximum metallicity is reached at $\sim 2$ Gyr,
([O/H]=1.3 and [Fe/H]=8.9 at $r_t$)
and then decreases due to the dilution effect of stellar mass loss,
as a higher fraction of H-rich gas is restored by lower-mass stars.
The present-day  metallicity of the gas is still high
([O/H]=1.1 and [Fe/H]=3.0 at $r_t$).
It is interesting to note that the oxygen gradient is steeper than the
iron gradient ([O/H]=2.8 and [Fe/H]=3.9 at 1 kpc).

 \begin{figure}
 \vspace{86mm}
 \caption{[Fe/H] and [O/H] radial profiles of the gas for the fiducial model 
at several epochs:
0.03 Gyr (solid line), 0.35 Gyr (short-dashed), 1 Gyr (long-dashed),
1.6 Gyr (dotted), 1.9 Gyr (dot-short-dashed), 4.4 Gyr (dot-long-dashed),
13 Gyr (short-dashed-long-dashed).}
 \end{figure}

 \begin{table*}
 \begin{minipage}{160mm}
  \caption{Present-day stellar chemical abundances}
  \begin{tabular}{@{}lccccc}
Model 
& $\langle$[Mg/H]$\rangle_1$ & $\langle$[Fe/H]$\rangle_1$ 
& $\langle$[Mg/H]$\rangle_{10}$ & $\langle$[Fe/H]$\rangle_{10}$ \\
 & & & & \\[3pt]
1 & 2.20 & 1.38  & 1.30 & 0.716 \\
2 & 2.57 & 1.53 & 1.38 & 0.754 \\
5 & 2.59 & 2.05 & 1.54 & 0.943 \\
10 & 3.40 & 2.22 & 1.58 & 0.935 \\
20 & 3.52 & 2.89 & 1.58 & 0.979 \\
50 & 4.30 & 3.98 & 1.69 & 1.420 \\
2(1/3) & 2.59 & 1.65 & 1.46 & 0.824 \\
2(0) & 1.57 & 0.66 & 0.77 & 0.283 \\
  \end{tabular}
\end{minipage}
\end{table*}

Since the gas removed with the wind represents a mass amount
comparable to the stellar component of the \gal,
it will have an important impact on  the intracluster or
intragroup medium where the \gal\ is located.
In particular, the \gw\ carries metals synthesized by the stars
and therefore it is of great importance
for the understanding of the chemical enrichment of the intergalactic
medium. 
The results of our models confirm the suggestion that elliptical galaxies
could explain the iron abundances in the intracluster medium (ICM) of
X-ray clusters of galaxies (David et al.  1990,1991). 
It is interesting to note that the late wind phase (i.e. from $t=1.60$ Gyr on)
makes the more important contribution to the iron enrichment. 
Only $6.9\times 10^7$ $\msun$ of iron is expelled during 
the early wind phase, that is, 14\% of all iron ejected by the \gal.
The oxygen enrichment of the intracluster medium occurs much earlier,
with $4.7\times 10^8$ \msun\ being expelled during the early wind phase,
i.e. half the total amount of oxygen ($9.83\times 10^8$ \msun)
ejected by the present epoch.
This result is not unexpected, since iron arises mostly from Type~Ia SNe and
the longer lifetimes of the  Type~Ia SN progenitors result in a delayed
Fe enrichment.
By contrast, the early wind removes gas pre-enriched by $\alpha$-elements
produced in Type II SNe during the first few $10^8$ yr of \ev.
In any case, one should take note of the important delays in the enrichment
of the ICM represented by the time scale $t_{w,e}$,
which is $\ga 1.5$ Gyr for \gals\ brighter than $\sim 0.1 L^*$,
and in the case of iron enrichment, by the arrival time of the Fe-rich
front at the boundary of the \gal\ (1.77 Gyr for the fiducial model).

The iron contribution to the ICM by the ellipticals could be checked through
the quantity $(M_{Fe}/L_B)_{cl}$, introduced by Ciotti et al. (1991)
(here, $M_{Fe}$ is the total iron mass in the ICM
and $L_B$ is the integrated blue luminosity of the \gals\ of the cluster).
Elliptical galaxies contribute most of the luminous mass in the most X-ray
luminous clusters. 
In the fiducial model, the \gw\ ejects until the present 
$4.93\times 10^8$ \msun\ of iron, which is mixed to the ICM.
Considering this model as representative of the cluster ellipticals, and
assuming $[M/L_B]=10$ for the stellar mass in the cluster,
from $M_*=2.38\times 10^{11}$ \msun\, we obtain
$(M_{Fe}/L_B)_{cl}=2.1\times 10^{-2} \msun/\lsun$
in good agreement with the value inferred from observations,
$(M_{Fe}/L_B)_{cl}\approx 10^{-2} \msun/\lsun$ (Ciotti et al. 1991),
therefore giving support to the thesis that enrichment by ellipticals
explains the ICM iron abundances.
The ratio $M_{w,Fe}/M_*$ is approximately constant $\approx 2\times 10^{-3}$
over the mass range $M_G=10^{11}-10^{12}$ \msun\ 
and also for the model 2(1/3).
In the most massive models 20 and 50, this ratio is reduced
to $1.4\times 10^{-3}$ and $7.5\times 10^{-4}$, respectively,
because the increase of the iron abundance as a result of  a longer building-up
time does not counterbalance
the decrease of the fraction of mass ejected
with the depth of the potential well.
In model 2(0), the higher $M_{w,Fe}/M_*$ ($3\times 10^{-3}$)
is due to the large amount of metals ejected with the wind,
with little metal being locked in stars.

We should take note of one possible problem regarding
the iron abundances of the wind predicted by our models, which
are too high in comparison to those derived from X-ray observations.
From BBXRT observations and one-temperatures fits, Serlemitsos et al (1993)
have estimated the metallicity of NGC 1399 and NGC 4472 to be
0.18-1.38, and 0.11-0.66 times solar  (90\% confidence), respectively.
This low abundance is in contradiction with $ROSAT$ measurements 
of Forman et al (1994), who determined
the iron abundance of NGC 4472 to be 1-2 times solar.
A more extended sample of early-type galaxies observed with the ASCA satellite,
however, has confirmed  the subsolar abundances found by the BBXRT.
The analysis of  galaxies in the Virgo cluster, including NGC 4742,
finds that the abundances of Fe and other elements are $\sim 0.5$ solar
(Awaki et al. 1994; Matsushita et al. 1994); 
and exceptionally low abundances, $Z\sim0.15$ solar,
are derived for NGC 1404 and NGC 4374 (M84) located
in the Fornax and Virgo clusters, respectively (Loewenstein et al. 1994).
However, the present X-ray spectral analysis has several uncertainties
that may lead to spurious results for abundances.
For instance, problems have been encountered with the basic plasma emission
models for the Fe L complex during the analysis of the ASCA spectra of the
cooling flows in the Perseus, Centaurus and A1795 clusters (Fabian et al. 1994).
The consistency of the spectroscopic data with the plasma emission models
was achieved only with a recalculation of the Fe L-shell spectra with a novel
atomic physics package and an update of the Fe data (Liedahl et al. 1995).

These low abundances, if real, severely constrain the Type Ia supernova
enrichment of the ISM of early-type galaxies.
In particular, the present SN Ia rate may be lower than that of our models.
In fact, the SN I rate predicted by the fiducial model
is at the upper end of the values derived from the observations;  
a reduction by 1/4-1/3 is still consistent with the observations.
On the other hand, 
Type II SNe may have a larger share in the Fe enrichment of the ICM,
either by an IMF flatter than the Salpeter one
as the $x=0.95$ IMF invoked by Arimoto \& Yoshii (1987) 
to reproduce the broad-band colors of giant \el s,
or by an increased lower-mass cutoff  of the IMF,
as in the high-mass mode of star formation, proposed
for the early \ev\ of \el s 
(Elbaz, Arnaud, \& Vangioni-Flam 1995).
Another possibility is that the abundances in the hot gas arise from
incomplete mixing of the supernova ejecta as relatively
metal rich inhomogeneities cool very fast decoupling from the hot gas
(Loewenstein et al. 1994).
Finally, in the case of galaxies in clusters, the infall of
gas from the ICM leads to metallicities
in the X-ray galactic corona
reflecting the subsolar abundances of the ICM.

\section{Abundances of the Stellar Population}

Table 2  shows the abundances of the stellar  population at the present:
$\langle$[Mg/H]$\rangle_1$ and $\langle$[Fe/H]$\rangle_1$,
the  Mg and Fe abundances mass-averaged over the inner kpc;
$\langle$[Mg/H]$\rangle_{10}$ and $\langle$[Fe/H]$\rangle_{10}$,
the  Mg and Fe abundances mass-averaged  inside 10 kpc
(10 kpc is roughly the effective radius of an $L^*$ \gal).
The code for the models is the same as in Table 1.
The core ($r\la 1$ kpc) of our model galaxies is metal-rich, in agreement with
the metallicities of 2-3 for the nucleus of \el\ \gals,
derived from observations of the Mg$_2$ indices.
For the fiducial model, 
both magnesium and iron exhibit central overabundances at the present 
([Mg/H]=3.4 and [Fe/H]=2.5 at 100 pc; [Mg/H]=2.0 and [Fe/H]=1.2 at 1 kpc).
At larger radii, however, the abundances become subsolar
([Mg/H]=0.67 and [Fe/H]=0.29 at 10 kpc),
indicating the presence of abundance  gradients in the \gal\ stellar population.

Metallicity gradients for the stellar population of early-type galaxies
have been inferred from measurements of line-strength gradients
(Gorgas, Efstathiou \& Salamanca 1990; 
Worthey, Faber \& Gonz\'ales 1992; Davies, Sadler \& Peletier 1993
Carollo, Danziger \& Buson 1993; Fisher, Franx \& Illigworth 1995). 
Fig. 7 shows the \ev\ of the magnesium and iron abundances of the stellar population
at several radii for the fiducial model, allowing us to derive metallicity gradients.
For $r>1$ kpc,
the abundance gradients predicted by our models are higher than
the average logarithmic abundance gradient ($d\log$[Fe/H]$/d\log r$) of -0.2 
derived by Davies et al. (1993) from line-strength gradients measured
out to distances of 5-20 kpc from the centre of the \gal.
For the magnesium, the present abundance gradients of the fiducial model
are -0.22 for 100 pc $<r<$ 1 kpc and -0.48 for 1 $<r<$ 10 kpc,
and for the iron, the respective values are -0.32 and -0.60.
The abundance gradients are expected to be flatter
when the dependence of the SFR on density is weaker. 
This is confirmed by the  model 2(1/3)
which exhibits magnesium (iron) abundance gradients of 
-0.18 (-0.28) for 100 pc $<r<$ 1 kpc and -0.49 (-0.61) for 1 $<r<$ 10 kpc,
although the abundance gradients do not change dramatically.
It should be noted that our models tend to overestimate the abundance gradients,
since the stars are moved only to reproduce a King profile:
the newly formed stars are stored in shells, and each shell is moved
as a whole entity to their final position.
Our models do not include orbital mixing, 
i.e., the distribution of orbits of individual stars,
in which case the stars at a given radius would have apocentres
spanning a wide range of radii.
The inclusion of orbital mixing would lead to the
flattening of the metallicity gradients.
This effect, however, would have little impact on the central abundances
of the galaxies. 

Note that a high metallicity stellar core is rapidly built up.
The central stellar magnesium
abundance becomes supersolar at $1.9\times 10^8$ yr 
and the iron abundance at $8.3\times 10^8$ yr, respectively.
Therefore, the conditions required by the starburst model of QSOs
(TB93), a metal-rich stellar population and
a metal-rich ambient gas (as seen in Section 4),
are achieved in the early stages of the \ev\ of \el s.

 \begin{figure}
 \vspace{86mm}
 \caption{The \ev\ of  [Mg/H] (solid curves) and [Fe/H] (dotted curves)
of the stellar population at several radii for the fiducial model. 
The upper, the middle and the lower
curves refer to radii of 100 pc, 1 kpc, and 10 kpc, respectively.
Note that the $r=100$ pc and $r=1$ kpc curves,
for both [Mg/H] and [Fe/H], {\it do not} cross.
}
 \end{figure}

One important constraint to \chev\ is given by abundance ratios derived from
Mg and Fe line strengths (based on Mg$_2$, Fe5270 and Fe5335 indices).
It seems that Mg is overabundant with respect to Fe in giant \el s.
The ratio [Mg/Fe] exhibits a large scatter around 1.5-2.0
(Worthey et al. 1992).
This high ratio is interpreted as a signature of enrichment by Type II SNe,
implying that the formation of an \el\ was very rapid, otherwise
the iron enrichment by Type Ia SNe would have shifted the [Mg/Fe] ratio
to solar or subsolar values.
Our models predict a range of $1.1-1.8$ for [Mg/Fe].
This reproduces well the observations,
indicating that the \sfor\ time scale of $10^8$ yr chosen
as normalization of the SFR based on results of one-zone models,
is also appropriate for the our dynamical model.
Again, model 2(0) is discrepant from the rest. 
As a consequence of the very early \gw\, 
[Mg/Fe] is shifted to very high values, $2.4-2.7$,
due to a very high dominance of SN II in the chemical enrichment,
and little metal is locked in the stars,
so the core is not metal-rich.

One important result we can derive from Table~2
is that the metallicity increases with the \gal\ mass,
and thus the well-known metallicity-mass relation for ellipticals is reproduced by our models.
The trend of increasing metallicity with mass is stronger for
the abundances averaged inside a  smaller aperture.
For a larger aperture, the metallicity-mass relation is less smooth. 
The  metallicity-mass relation is a consequence of the occurrence of  \gw s,
which appear later for \gals\ with deeper gravitational wells,
thus allowing more metal enrichment for the gas and the stars formed from it,
a mechanism first suggested by Larson (1974b).
The consistency of the \gw\ scenario can be seen from the results of
Table 1, which shows that, given a star formation law,
the time for the onset of the \gw\ increases with the galaxy mass.

One further prediction from Table 2
is a clear, although not strictly monotonic, trend
of  decreasing [Mg/Fe] ratio with the galactic mass:
for the models with standard SFR, 
$\langle$[Mg/Fe]$\rangle_1$ ($\langle$[Mg/Fe]$\rangle_{10}$)
shows a maximum of  1.68 (1.83) for model 2 
(note that the respective values for model 1 are 1.59 and 1.82)
and a minimum of 1.08 (1.19) for model 50.
However, this prediction seems to be at variance with the observations 
of the stellar population in \el s,
which suggest a slight trend of increasing [Mg/Fe] ratio
with the galactic luminosity, albeit with a large scatter
(Worthey et al. 1992, Weiss, Peletier \& Matteucci 1995).
The reason for the increase of  [Mg/Fe] with $M_G$ in our models
is that, under the our assumption of  star formation time scale independent of
galactic mass ($\nu_0^{-1}=10^8$ yr), 
the \gw\ happens later for more massive objects.
As a result, the [Mg/Fe] ratio tends to decrease with the galactic mass because
the bulk of iron is produced by type Ia SNe with a $\sim 1$ Gyr delay relative to
the $\alpha$-elements, produced by type II SNe on much shorter time scales.
Note that, for the sake of simplicity, our models assume that:
1) the specific SFR is independent of the galactic mass; 
2) the IMF is the same for all \el\ \gals\ (see Matteucci 1994).
If we relax the assumption 1), 
the predicted [Mg/Fe] trend could be conciliated with the observations
if the \sfor\ in giants \el s is faster than in smaller \el s, 
so favouring magnesium production over iron production.
Another way to achieve a increase of [Mg/Fe] with the galactic mass is
a variable IMF, with the slope of the IMF decreasing with the galactic mass,
which leads to a larger proportion of massive stars in larger galaxies,
and, as a consequence, to a larger $\alpha$-element enhancement in more massive \gals.

\section{Evolution of the Luminosity of QSOs}

\begin{table*}
\begin{minipage}{160mm}
\caption{Central cooling flow properties}
\begin{tabular}{@{}lccccccc}
  Model & $M_{c}(t_G)$ & $N_{cf}$ & $t_{cf,1}$ & $M_{c,1}$ 
& $M_c^*$ & $\zeta$ & $n_{c,1}$ \\
& (10$^8$ $\msun$) & & (Gyr) & (10$^8$ $\msun$) 
& (10$^8$ $\msun$) & ($10^{-3}$) & \\[3pt]
1 & 5.20 & 1 & 1.80 & 5.20 & 4.15 & 3.94 & -2.53 \\
2 & 6.74 & 1 & 1.80 & 6.74 & 5.53 & 2.83 & -3.58 \\
5 & 13.2 & 1 & 2.23 & 13.2 & 9.72 & 2.24 & -3.23 \\
10 & 17.9 & 2  & 2.81 & 17.4 & 12.6 & 1.50 & -4.25 \\
20 & 51.7 & 2 & 2.78 & 35.2 & 24.3 & 1.52 & -4.28\\
50 & 185  & 2 & 3.16 & 97.1 & 67.4 & 1.46 & -3.68 \\
2(1/3) & 8.48 & 3 & 2.04 & 8.24 & 6.05 & 2.91 & -3.02 \\
2(0) & 1.76 & 1 & 1.36 & 1.76 & 1.62 & 1.32 & -2.16 \\
\end{tabular}
\end{minipage}
\end{table*}

The luminosity function (LF) of QSOs undergoes
strong \ev\ between $z=2$ and the present epoch, with the redshift
dependence of the LF being well-described by a constant comoving space
density  and a pure power-law luminosity \ev\ $L(z) \propto (1+z)^k$
(Boyle et al. 1988, 1991, BT98).
Estimates of the rate of \ev\ $k$ lie in the range $3.1<k<3.6$,
corresponding to an \ev\ in terms of cosmic time $t$ of the form
$L_{QSO}(t) \propto t^{-n}$, 
with $2 < n < 3.5$, depending on the choice of cosmology.

Both in the supermassive black hole scenario and
in the starburst model for AGN, the QSO luminosity \ev\ is linked to 
the mass flow into the galactic nucleus.
In both cases the QSO luminosity is proportional to the average mass accretion rate,
via either gravitational energy (supermassive black hole)
or massive stars activity (starburst model).
Two central assumptions are common to both approaches 
to explain the luminosity function
of QSOs: 1) the luminosity of the QSO is proportional to the mass of the 
host galaxy; and 2)  virtually all bright \gals\ have harboured a QSO (AGN)
during their \ev. 

Our models develop a central \cf\ as soon as
the gas central density has increased enough to allow radiative losses
to be important.
The \cf\ extends over a considerable span of their \ev,
and, in most cases, when enough gas has been consumed by \sfor\ or removed by gas flows,
SN heating becomes more important than
radiative cooling, and the central \cf\ is extinguished.
As shown below, 
the fate of the central \cf\ depends on the galaxy mass and the \sfor\ law.
We now proceed to investigate whether the time-dependency of the 
central \cf\ rate in our models could be responsible for the observed
evolution of QSOs.

Table 3 gives relevant information about the \ev\ of the central \cf\ for 
several models (see Table 1 for the keys to the models):
$M_{c}(t_G)$, the mass deposited by the 
central (into the inner 100 pc) \cf\  until the present;
$N_{cf}$, the number of \cf\ episodes until the present;
$t_{cf,1}$ and $M_{c,1}$, the duration and
the total mass deposited by the first \cf\ episode;
$M_c^*$,  the mass deposited  by the \cf\ until $t=1$ Gyr;
$\zeta=M_{c,1}/M_G(t_G)$ (this parameter relates the present-day \gal\ mass
to the mass of galactic nucleus at high-redshifts).
The end of the first central \cf\ episode is defined 
when the central  \cf\ rate $\dot M_c=0$
or, if  the central \cf\ never vanishes, when $\dot M_c$
reaches a minimum.

Fig. 8 shows the \ev\ of $\dot M_c$.
In order to compare to the observed LF \ev\ 
(Boyle et al. 1988, 1991, BT98), we fitted a power law to $\dot M_c$
between $t=1$ Gyr and $t_{cf,1}$.
The values of the index $n_{c,1}$ of the fit $\dot M_c \propto t^{n_{c,1}}$
 are shown in Table 3.
As we can see from $M_c^*$ (Table 3), $\approx 70-80$ \% of the total mass
of the first central inflow have been deposited in the galactic nucleus by 1 Gyr.
This time is the typical time scale  for the completion of the large spheroids.
If we consider that this event concides with the QSO phenomenon,
the epoch when the QSO activity was at its peak ($2<z<3$) (Schmidt et al. 1991)
corresponds to  the epoch of galaxy formation $3.21 < z_{GF} < 6.54$ Gyr,
a reasonable range in a number of scenarios for \gal\ formation.

As seen in Section 3,
the initial global inflow stage is followed by a partial wind stage,
during which a wind is established in the outer regions of the galaxy.
The stagnation region separating the wind and the inflow moves inwards
until only a wind is present throughout the galaxy.
In some cases, the total wind persists until the present epoch and no material
restored by the stars is accumulated in the galaxy. 
That is the {\it single partial wind} sequence (the partial wind occurs only once),
\vskip 2mm
global inflow $\rightarrow$ partial wind $\rightarrow$ total wind.
\vskip 2mm\noindent
In some models, however, after some time, the gas in the inner regions
is not swept out rapidly enough and starts accumulating until a central
cooling flow is established again. 
The occurrence of the late partial wind characterizes a second type of gas flow sequence,
the {\it multiple partial wind} sequence
\vskip 2mm
global inflow $\rightarrow$ partial wind $\rightarrow$ total wind
$\rightarrow$ partial wind...
\vskip 2mm\noindent
The three models which exhibit late partial winds 
--- models 10, 20, and 2(1/3) --- illustrate
the three possible fates of the late partial wind.
In the first place, the partial wind could persist until the present (model 10).
Other possibility is that it develops into a global \cf\ extending from the nucleus
to the tidal radius of the \gal.
That is the case of model 20, in which the outer wind reverses to outer inflow
at 11.8 Gyr.
Finally, the central inflow could be halted by supernova heating due to the
resulting starburst occurring in the \gal\ core, as in model 2(1/3).
There is one further type of gas flow sequence, in which the total wind
never occurs, that is,
\vskip 2mm
global inflow $\rightarrow$ partial wind $\rightarrow$ global inflow.
\vskip 2mm\noindent
This sequence is exhibited by the very massive model  50.
In this model, the central \cf\ is never turned off, and more than
$10^{10}$ $\msun$ has been deposited
into the nucleus by the present epoch.
Due to the large potential well of the \gal, the \gw\ stalls early and an
outer inflow is established at 8.8 Gyr.

As seen from Fig. 8,
in our models, the central inflow rate exhibits
a large variety of evolutionary sequences,
some of which combine
features of  the continuous and recurrent activity,
exhibiting in many cases late \cf\ episodes, which can be
weak or vigorous, short- or long-lived.
All the models, however, show during
the first Gyr a massive deposition rate,
followed by a declining accretion rate \ev.
The first  central \cf\ episode, common to all models,
 extends for $2-3$ Gyr.
The early \ev\ of the first \cf\ is subject to considerable fluctuations. 
An early, secondary maximum is apparent between 30 and 60 Myr,
and it is followed by a decrement in the inflow rate, as the heating
due to Type II SNe inhibits the \cf.
From $10^8$ yr on, the inflow rate increases until it reaches the absolute maximum.
By then, and during the following decline of $\dot M_c$, the \ev\ is relatively smooth.
The peak inflow rate occurs at $0.38-067$ Gyr for the models with $n_{SF}=1/2$,
 and is earlier for the most massive models.
For model 2(1/3), it is at 0.94 Gyr, and for model 2(0), it is at 0.17 Gyr.

The fit to the declining part of the \ev\ of the \cf\ rate
by $\dot M_c \propto t^{n_{c,1}}$
gives values of $n_{c,1}$ in the range $-4.3< n_{c,1}<-2.5$.
The values of $n_{c,1}$ refers only to the first \cf\ episode
and describe only the 1-3 Gyrs following the maximum in  $\dot M_c$.
The principal conclusion that $n_{c,1}$ (and $t_{cf,1}$) allows us to derive
is that, under the assumption of proportionality of the luminosity to $\dot M_c$,
the most active phase of the QSO is likely to occur during a short span
of time since the \gal\ formation and that
the early \ev\ of luminosity is very steep,
which is consistent with the derived \ev\ of the QSO LF.

One particularly interesting model is 2(1/3)
because it exhibits recurrence of late central inflow episodes.
As the SN I rate decreases, the wind stalls, the gas starts accumulating
in the \gal\ core, and the radiative losses drive a new central \cf.
The late central inflow episodes are brief
and quickly put out by the resulting burst of star formation.
The duration of the late \cf\ episodes is a few times $10^7$ yr,
and, since the separation between them is typically a few Gyr,
their duty cycle is $\sim 1/100$.
The turning on and off of the central inflow could correspond to the
``active" and ``inactive" states of the core in
both the supermassive black hole model of AGN
and the starburst model of AGN.
In addition, the length of the duty cycle of the central inflow
increases with time and the amount of material deposited
decreases, implying a steep decrease of the time-averaged
luminosity of any central activity powered by the infalling material.
For the model 2(1/3),  the second central inflow occurs 
at $t=4$ Gyr and involves $1.5\times 10^7$ $\msun$, 
and the third one of $8.4\times 10^6$ $\msun$ happens at $t=9.5$ Gyr.
Note that also in model 20, the late central inflow episode 
at $t \sim 5$ Gyr is inhibited by SN heating,
following the starburst in the galactic core.
However, the deep potential well of this model guarantees that at the end
a massive central inflow prevails.

Episodic models of QSO activity have already been put forward
to explain the observed \ev\ of the QSO LF
both in the supermassive black hole scenario (Cavaliere \& Padovani 1986)
and in the starburst model (TB93).
In the starburst model, the SFR in the galactic core actually responsible
for  the AGN activity is represented by a series of narrow ($\sim 10^7$ yr) peaks.
In this model, between $z=2$ and the present,
the galaxy core undergoes between 3 and 5 successive bursts of QSO activity,
with their amplitude being modulated by a power law with time of the form
$SFR_{max} \propto t^{-2}$.
This power law \ev\ is found in the elliptical galaxy models of Larson (1974a),
which, however, refers to a continuous \ev\ of the SFR instead of an episodic one.

It is interesting that just changing the \sfor\ law
from $n_{SF}=1/2$ in the fiducial model to $n_{SF}=1/3$
gives rise to the appearance
of short episodes of inflow, lasting $\sim 3\times 10^7$ yr,
similar to those appearing in the episodic scenario.
As matter of fact, the time scale above refers to the final, more vigorous part
of  the inflow episode, when the central inflow rate suddenly becomes
larger than $10^{-2}-0.1$ $\msun$ yr$^{-1}$.
The total duration of each episode is some $10^8$ yr,
characterized in its first phase by an inflow at a low rate
(cf. the left wing in the second \cf\ event of model 2(1/3) in Fig. 8).
Once the inflow rate becomes larger, the \cf\ is extinguished by the
resulting \sfor\ burst.
The fact that \ev\ of the inflow rate with recurrent late \cf\ episodes
appears naturally in model 2(1/3)
gives support to the episodic scenario of QSO activity.

The models with a deep potential well (models 10, 20 and 50)
have  massive \cf s that are well-developed at the present, 
in some cases having been initiated very recently (model 10).
In models 20 and 50, the late \cf \ is particularly large, 
with $1.3\times 10^9$ $\msun$ and $8.8\times 10^9$ $\msun$, respectively,
accumulated in the nucleus during the second central inflow episode.
These massive late \cf s are distinct from the short-lived events
required by the episodic models of QSO activity,
in that their duration is $\sim 1$ Gyr or more.

The very massive model 50 allows us to investigate the ability of
our models to account for the luminosities of the brightest QSOs.
The present stellar mass ($6.6\times 10^{12}$ \msun) of this model
is representative
of the most luminous ($M_B \approx -24$) ellipticals in the nearby Universe
(corresponding to a stellar mass of $6.2\times 10^{12}$ \msun\ for
$[M/L_B]=10$).
Model 50 has a peak central inflow rate of 20 $\msun$ yr$^{-1}$ 
at $3.8\times 10^8$ yr. 
The expected bolometric luminosity 
$L_{Bol}=f \dot M c^2
= 5.7\times 10^{46} f (\dot M/\msun\,{\rm yr}^{-1})$ erg s$^{-1}$,
is $1.1\times 10^{47}$ erg s$^{-1}$,
for an efficiency $f$ of mass-energy conversion of 0.1.
This bolometric luminosity is typical of very bright distant QSOs,
but it fails to explain the bolometric luminosities
of the brightest QSOs ($\approx 10^{48}$ erg s$^{-1}$).
It should be noted, however, that the
bright end of the QSO LF is contaminated by \lum\ overestimates
due to beaming and gravitational lensing.
In addition, the luminosity  estimated above
assumes that the mass flow rate through the inner 100 pc radius
is directly deposited into the central energy source of the QSO.
It is expected that the \ev\ of the central inflow rate will modulate any central activity
inside the galaxy nucleus, but the present calculations do not intend
to describe in detail the region inside the 100 pc radius.
As a matter of fact, the interaction gas flow-central engine may take
place at a much smaller scale than the inner boundary.
In particular, the bolometric luminosity given above refers to a continuous 
deposition of gas fueling the \lum.
The gas may as well accumulate in the core during a long period
and then be consumed in a much shorter ``turning-on" time, 
either falling into a supermassive black hole or triggering a violent starburst.
One useful time scale to constrain the turning-on time
is the crossing time though $r_{in}=100$ pc, $t_{cross}=r_{in}/\sigma$.
From the Faber-Jackson relation, $t_{cross}=5\times 10^5$ yr for an $L^*$ \gal,
and $2.5\times 10^5$ yr for a \gal\ with $M_B=-24$.
If the turning-on time is $10^6-10^7$ yr, and the accumulation time 1 Gyr,
the duty cycle is $10^{-3}-10^{-2}$,
a range expected in several models for QSO activity.
By the time of peak deposition rate in model 50, 
the mass in the nucleus is $2.8\times 10^9$ $\msun$.
By $t=1$ Gyr, it   is $6.7\times 10^9$ $\msun$.
A burst of activity involving such amount of mass
at a conversion efficiency $f=0.1$ with a time scale of a few $10^7$ yr,
could explain the bolometric luminosity of even the most luminous QSOs.
As a matter of fact, the limit luminosity to be attained by any sort
of central engine inside the galaxy nucleus is the Eddington luminosity
$L_E=4\pi G M m_H c/\sigma_T
=1.4\times 10^{38}\,(M/ \msun)$ erg s$^{-1}$.
For model 50 at $t=1$ Gyr, $M_c=6.7\times 10^9$ $\msun$ implies
$L_E= 9.4\times 10^{47}$ erg s$^{-1}$, a value that,
even disregarding beaming effects,  could explain the highest luminosities
of QSOs.
In addition if we take into account the trend of  $[M/L_B]$ to increase with
\gal\ mass (Terlevich 1992), then
the most luminous galaxies have $[M/L_B] \approx 20$ 
implying a larger mass than that inferred from the blue luminosity
using  $[M/L_B] =10$ . 
In this case, scaling the luminosities to model 50, 
$L_E$ exceeds $10^{48}$ erg s$^{-1}$.

 \begin{figure}
 \vspace{86mm}
 \caption{Predicted evolution of the central inflow rate $\dot M_c$
(binned over 20 logarithmic time intervals per decade). 
The panels are labelled according to the model codes of Table 1. 
The dashed lines represent power-law fits to $\dot M_c$
between $t=1$ Gyr and the end of the first \cf\ episode.}
 \end{figure}

The \ev\ of $M_c$ in our models allows us to make predictions about the
QSO LF.
Specifically, we now proceed to calculate the QSO LF in the redshift range
$2.0 < z < 2.9$ in order to compare our results with the observational
blue QSO LF of Boyle et al. (1991).
In view of the energetic dificulties with the continuous \ev\ of activity,
we assume that the \lum\ of the QSO is due to short episodes of
nuclear activity resulting from the fast consumption of the gas accumulated
in the nucleus by the central inflow.
For illustrative purposes we consider only the first inflow episode and
that two nuclear activity events occur during the first central inflow.
Based on the results of Table 3, we consider that the duration
of the first inflow is 2 Gyr. 
We assume the first activity event at 1 Gyr and the second one at 2 Gyr.
From the ratio $M_c^*/M_{c,1}$, we assume that 75 \% of $M_{c,1}$
is consumed by the first event and 25 \% by the second one.
For the estimate of the LF of galactic nuclei at high redshifts,
we need to scale the present-day \el\ \gal\ LF
both in \lum\ and in space density.
The present-day \el\ \gal\ LF we adopt is (Terlevich 1992; TB93)
\begin{equation}
\Phi(M_B)dM_B=\Phi^*  10^{0.4(M_B^*-M_B)\beta}
\exp [-10^{0.4(M_B^*-M_B)}] \; ,
\end{equation}
where $\Phi^*=3.6\times 10^{-4} \, h_{50}^3$ Mpc$^{-3}$,
$M_B^*=-21$, and $\beta=0.23$. 
This form represents well the LF of moderately
bright \el s, but underestimates the LF both at the bright and the faint ends.
Accordingly, a power-law extension for $M_B \leq -22.5$,
$\Phi(M_B)dM_B \propto (L_B/L_B^*)^{-3}$, is included
to represent the cD \gal\ data (Terlevich 1992).
One the other hand, recent determinations of the faint-end of the LF indicate
a turn-up of the LF for luminosities fainter than $M_B=-17.5$
(Smith, Driver, Phillipps 1997).
Therefore, following Smith et al., we add a dwarf Schechter LF
with parameters $\Phi^*$(dwarfs)$=1.5\times \Phi^*$(giants), $M_B^*=-17.5$,
and $\beta=-0.7$, to represent the dwarf spheroidals.

The scaling of the \lum\ $L_B^k$ of the k-th nuclear activity event 
to the present-day \el\ \gal\ luminosity $L_B$ is given by
\begin{eqnarray}
L_B^k &=&(f/BC)\xi_k (M_c c^2/t_{on})
 \nonumber \\
 &=&(f/BC)\xi_k (c^2/t_{on}) \times \zeta \times [M/L_B] \times L_B,
\end{eqnarray}
where  $BC=L_{Bol}/L_B$ is the nucleus bolometric correction factor,
$f$ is the efficiency of mass-energy conversion,
$M_c$ is the mass deposited by the first central inflow episode,
$\xi_k$ is the fraction of $M_c$ consumed in the k-th activity event,
$t_{on}$ is the time-scale for gas consumption,
$\zeta=M_c/M_*$, the ratio of $M_c$ 
to the present-day \gal\ stellar mass $M_*$,
and $[M/L_B]$, the present-day \el\ mass-to-light ratio.
The adopted $\zeta-M_*$ relation is derived from Table 3:
 $\zeta=3.28\times 10^{-2} -2.59\times 10^{-3}\log(M_*)$ 
for $M_* < 1.2\times 10^{12}$ \msun, 
and $\zeta=1.5\times 10^{-3}$  for larger masses.
The increase of $[M/L_B]$ with $|M_B|$
is represented by $[M/L_B]=-1.90-0.138 M_B$ (Terlevich 1992)
for $-18.8 \geq M_B \geq -23.2$, with $[M/L_B]=20$ for higher luminosities
and $[M/L_B]=5$ for lower luminosities.

In order to scale in density number we assume that the \gals\ are
formed uniformly in time during the epoch $3<z_{GF}<10$.
Considering that the high-redshift QSO activity
is signalled by the first nuclear activity event which occurs when the \gal\ is
1 Gyr old, the correponding epoch of nuclear activity is $1.86<z<3.52$.
This span in redshift is consistent is the strong decline of QSO counts
for $z<2$ and $z>3.5$ (Boyle et al. 1991, Warren et al. 1994).
The scaling in density is then obtained by multiplying the present-day
\el\ \gal\ LF by the duty cycle $\kappa=t_{on}/t_{off}$
($t_{off}=t(z=3)-t(z=10)=1.33$ Gyr).

 \begin{figure}
 \vspace{86mm}
 \caption{Predicted galactic nuclear $2<z<2.9$ LF within both
the massive black hole (dotted line) and the starburst (dashed line) scenarios.
Also shown (dot-dashed line)
the \ev\ of the LF predicted by a formal best fit model (see text). 
between $2.00 < z < 2.90$ and $1.25<z<2.00$.
The LF data of Boyle et al. (1991) 
are also given for the same redshift intervals
(symbols connected by solid lines).}
 \end{figure}

Fig. 9 illustrate the predictions of our models for the 
$2<z<2.9$ LF of the galaxy nucleus,
combining the two activity events described above,
for both the massive black hole and the starburst scenarios.
The relation $L_B^k - L_B$ can be rewritten in a convenient way as
\begin{equation}
L_B^k/L_B=14.7 \, (f_{0.1}/BC_{10}) \xi_k  (\zeta_{-3} /t_{on,7}) [M/L_B],
\end{equation}
where $f_{0.1}=f/0.1$,  $BC_{10}=BC/10$, 
$\zeta_{-3}=\zeta/10^{-3}$, and $t_{on,7}=(t_{on}/10^7 \, {\rm yr})$.
In the calculation of $t_{on}$, we assume that 
$t_{on} \propto t_{cross}$ or  $t_{on}= t_{on}^*(L_B/L_B^*)^{-1/4}$.
In addition, $t_{on}$ is constrained by the Eddington time
$t_E=\sigma_T c/4\pi G m_H$
($t_{on} \geq f\,t_E= 4\times 10^7 f_{0.1}$ yr)
and, in the starburst model, by the lifetime of the typical star giving
the \lum\ of the starburst ($t_{on} \geq t_m(\langle m \rangle)$).
For the massive black hole model, we consider a rather
conventional set of values $f_{0.1}=1$, $BC_{10}=1$, and $t_{on,7}^*=3$.
The values $f_{0.1}=0.05$, $BC_{10}=0.6$, and $t_{on,7}^*=1$ are used
in the starburst model.
We consider that the starburst is dominated by very massive
stars,  e.g. $\langle m \rangle \approx 25$ \msun,
since the standard star cluster
of TB93 model would not be energetically feasible.
From the fact that, for a $25$ \msun\ star,
$\sim 17$ \msun\ have\ been converted  into helium and carbon
at the end of the C-burning phase with a convertion efficiency of 0.007,
we obtain $f=(17/25)\times 0.007=0.005$.
The value $BC=6$,
although low,  has been previously considered in calculations of the QSO LF
within both the massive black hole and the starburst scenarios
(Haehnelt \& Rees 1993, Terlevich 1994).
The starburst considered in the present model differs from
that in TB93 model in that here, we focus
only on the galactic nucleus
(with $0.15-0.5$ \% of the \gal\ mass and $r<100$ pc).
In contrast to the {\it nuclear starburst} in this paper,
the {\it core starburst} in TB93 involves the whole core of the \gal,
comprising 5 \% of the \gal\ mass inside $300-2000$ pc.
As we can see from Fig. 9, the supermassive black hole model
leaves room for an efficiency of mass-energy conversion 
somewhat lower than $f=0.1$,
whereas the nuclear starburst, although it can explain the
luminosities of the brightest QSOs, 
systematically underestimates the density  number of QSOs,
due to the smallness of its duty cycle.
Note that in both cases, there is a remarkable agreement in shape between
the predicted and the observed QSO LF at high redshifts.
Also the early \ev\ of the LF is well described by our models.
This is illustrated in Fig. 9 by a formal best fit model
($f_{0.1}=0.3$, $BC_{10}=1$, and $t_{on,7}^*=1$)
 which adjusts the $2<z<2.9$ data.
In this case, the \ev\ from $2<z<2.9$ to $1.25<z<2$ is reproduced.
Since we are focusing only on the first inflow episode, which extends
only for $\sim 3$ Gyr,
the predictions of our models refer only to 
the early ($z>1$) \ev\ of the QSO \lum.

Note that the masses of the central black hole or nuclear star cluster
produced by the first inflow episode
(up to $\approx 10^{10}$ \msun) are consistent the evidence
of Dark Massive Objects (DMOs) in the nuclei of  both active and quiescent \gals\ 
(see reviews by Kormendy \& Richstone 1995, and van der Marel 1996).
So far, the masses are in the range $2\times 10^6-3\times 10^9$ \msun,
with the largest mass found for M87.
There is a correlation of increasing MDO mass with the luminosity of the \gal,
and, since M87 has $M_B=-21.4$,
DMOs as massive as $10^{10}$ \msun\ are expected for $M_B=-24$ \gals.
But such \gal s are very rare, and the corresponding DMOs would be unlikely
to be found in the volume of the Universe searched for DMOs until now
(M87 is the farthest DMO host, at 15 Mpc).
Finally,
one important success of our models is
that the average MDO-to-galaxy mass ratio of $\sim 0.003$
(Kormendy \& Richstone 1995)
is reproduced by  the range $\zeta =0.0015-0.005$
of our models.

\section{Conclusions}

In this work, we explored the relation between young \el\ \gals\ and QSOs
within a chemo-dynamical model for \ev\ of \gals.
In this model, we perform a multi-zone modelling
of  the chemical enrichment of the gas and stars of a  \gal\ 
taking into account the gas flow
obtained self-consistently from 1-D hydrodynamical calculations.
We were particularly interested in the \cf\ towards the centre of the \gal,
which could feed a QSO hosted in the galactic nucleus.

From a minimal set of assumptions,
based on standard one-zone \chev\ models,
our model reproduces the main observational features of \el\ \gals:
1) the central metallicities of massive \gals\ are supersolar;
2) the ratio [Mg/Fe] is supersolar
 in the core of the \gal\ as well as over the whole \gal;
3) the \gal\ shows sizable metallicity gradients;
4) systems with larger masses tend to have larger metallicities,
thus reproducing the mass-metallicity relation of \el\ \gals; 
5) elliptical galaxies can be the main responsibles
for the ICM iron content: the ICM iron mass per unit luminosity
of cluster galaxies ($\sim 10^{-2}$ \msun/\lsun) is reproduced.

One very important time scale for the enrichment of the ICM is
the time $t_{w,e}$ of the end of early wind phase, during which the gas content
in the \gal\ is reduced by a factor of ten.
$t_{w,e}$ is longer than $\sim 1.5$ Gyr for normal or bright \gals.
For the iron enrichment, the relevant time scale is even longer,
because the Fe-rich gas front arrives at the galaxy edge  later  than $t_{w,e}$.

From the success of the model in reproducing \el\ \gals,
a number of standard assumptions of the classic one-zone chemical \ev\ models
remain valid when the hydrodynamics is considered:
$10^8$ yr \sfor\ time scale, Salpeter IMF, stellar yields, $A_{SN\,I}=0.1$.
However,
the linear \sfor\ law usually adopted by one-zone models
seems to be excluded, since the resulting model
would be unrealistic, not reproducing the properties of \el\ \gals,
as seen from
the drawbacks of model 2(0):  too low central metallicities;
too high [Mg/Fe] ratio; the nucleus mass is too small to explain
the luminosity of a QSO.

We should take note of two possible discrepancies
between the results of our models and the observations:
1) supersolar metallicities are predicted
for the hot gas in the galactic halo, 
while ASCA measurements imply subsolar abundances;
2) we predict a trend of [Mg/Fe] decreasing with galactic mass,
which is the opposite of the trend inferred from determinations of metal indices.
If these contradictions are real, they could indicate that
some of the standard assumptions of the \chev\ modelling do not apply.
We should be aware, however, of the uncertainties in deriving the abundances
in the hot galactic halos (besides the dilution by the ICM, which lowers
the metallicity in the halo),
and of the difficulties in obtaining [Mg/Fe] trends from the observed variation
of Mg and Fe line strengths with the galaxy mass.

In addition to the fact that the model satisfy a number of observational
constraints on \el s, it makes definite predictions about the relation
between the \ev\ of \el s and that of QSOs.

A high-metallicity core is rapidly built-up. For the gas, 
solar metallicities are reached in $10^8$ yr for oxygen, and $3\times 10^8$ yr
for iron. For the stars, the magnesium abundance becomes solar
at $2\times 10^8$ yr and the iron at $8\times 10^8$ yr.
In this way, the high abundances derived for high redshift QSOs are reached
in a reasonably short time scale.
One important application of these results is that
the enrichment time scales predicted by our chemo-dynamical
model provide a chemical clock which could constrain cosmological
scenarios. For instance the $\sim 1$ Gyr time scale for metal production
implies an age at least larger than 1 Gyr for the Universe at $z \sim 5$,
since even the more distant QSOs exhibit metals in their spectra.
In addition, the short time scales for enrichment both of gas and stars
make plausible the starburst model for AGN, which requires
a high metallicity core formed early in the \ev\ of the \gal.

It is important to note that
the luminosity and metallicity of QSOs identified with the central
region of the young elliptical are explained with no need for all
the galaxy having a global starburst coordinated with the central starburst.
In this way, extremely high luminosities ($\sim 10^{15}$ \lsun)
are avoided for the proto-QSO.
These extreme luminosities, predicted by one-zone models of 
formation of \el\ \gals\ (e.g. model M4 of HF93),
have never been detected in high redshift observations.
Note that probably only the inner regions of the young \gal, 
due to their higher surface brightness,
would have \sfor\ detectable at high redshifts.
As a matter of fact, deep spectroscopy observations have 
revealed a population of star-forming \gals\ at redshift $3 \la z \la 3.5$,
the Lyman break \gals\ (LBGs),
discovered on the basis of a Lyman limit break 
superposed on their UV continuum 
(Steidel et al. 1996; Giavalisco, Steidel \& Macchetto 1996).
The LBGs show many characteristics expected for primeval \gals,
and, assuming a Salpeter IMF, their inferred SFRs are in the range
$4-90 h_{50}^{-2}$ \msun\ yr$^{-1}$, inside a typical half-light radius
of $1.8 h_{50}^{-1}$ kpc ($q_0=0.5$).
These levels of \sfor\ are remarkably close to 
the typical range of $20-50$ \msun\ yr$^{-1}$
for the SFR inside a radius of 1 kpc in the fiducial model.
In addition, the LBGs are expected to be the high-redshift
counterparts of the present-day spheroidal component of luminous \gals,
since their co-moving density is roughly comparable to that of
present-day bright ($L \geq L^*$ )\gals\ and 
the widths of the interstellar absorption lines
in their spectra imply circular velocities of $170-300$ km s$^{-1}$,
typical of the potential well depth of luminous \el s.
Therefore, for $z \la 3.5$, the observations of
ongoing \sfor\ in \gals\ seems to rule out the high luminosities
predicted by the one-zone models for the progenitors of
present-day $\sim L^*$ \gals.

The luminosities of the one-zone model could still be consistent
with the observations of LBGs, if there is dust absorbing the blue and
UV light and re-emitting it in the far-IR/sub-mm.
However, comparisons between the observed colours of LBGs
and those predicted by spectral synthesis models
(Pettini et al. 1998) indicate only modest dust attenuation 
(an extinction at 1500 \AA\ between
lower and upper limits of $\sim 2$ and $\sim 6$), 
and, therefore, dust absorption is unable to hide 
the high luminosities of the one-zone model.
In addition, 
in view of the evidence that the LGBs are the progenitors of 
the present-day luminous ($\sim L^*$ or brighter) \gals, 
rather than sub-units being assembled into a larger system,
it seems that the LBGs are allowing us to witness one stage
in a single event of formation of a massive \gal,
with the global star formation proceeding, however, at a milder rate
than in the one-zone model.
In the end, 
the main reason why the one-zone model overpredicts the luminosity
is that it overproduces metals.
The model M4 of HF93 predicts $\sim 10$ times the solar metallicity
over the whole \gal.
The observations, however, allow
this extremely high metallicity only at the very nucleus of the \gal, at most.
Over an effective radius, the metallicity of a giant \el\ is roughly solar.
These metallicities are correctly predicted by our multi-zone model
(see the values of  
$\langle$[Mg/H]$\rangle_{10}$ and $\langle$[Fe/H]$\rangle_{10}$ in Table 3).
Moreover, assuming
that the observed negative metallicity gradients continue
beyond the effective radius,
the mass-averaged metallicity would be subsolar over the whole \gal.
Therefore, scaling the metals produced in the one-zone approximation
to amounts consistent with the observations,
brings down the predicted luminosities by about one order of magnitude.

All our models predict a massive central (through the inner 100 pc) \cf\ 
during the first 1-2 Gyr of the \gal\ \ev.
Within the scenario in which the luminosity of the galactic nucleus is fed by
the central inflow,
for a reasonable epoch of formation of the spheroidal systems,
the epoch of building-up of the nucleus by the central inflow
coincides with the maximum in QSO activity ($2<z<3$).
Also the decrease of the central inflow rate for $t>1$ Gyr
is consistent with the decline of luminosity inferred for $z<2$ from the
\ev\ of QSO LF with redshift.

One of our models exhibits  recurrent late central \cf\ episodes,
which are brief (a few $10^7$ yr) and involve decreasing amounts of mass.
The gas inflow into the inner 100 pc is regulated by episodes of
star formation
leading naturally to several short episodes of central inflow,
thus
giving support  to the episodic scenario for evolution of the LF of QSOs.

The model also explains the luminosities of QSOs.
The central \cf\ rates explain bolometric luminosities of
up to $10^{47}$ erg s$^{-1}$,
for an efficiency $f$ of mass-energy conversion of 0.1.
However, the bolometric luminosities of the brightest QSOs
($\approx 10^{48}$ erg s$^{-1}$)
cannot be explained by a continuous deposition of the central inflow.
Rather, the highest QSO luminosities require a discrete deposition,
in which the gas is  accumulated
during $\approx 1$ Gyr and then consumed by a central engine in few $10^7$ yr.
Accordingly, 
we have made some predictions on the QSO LF at $z \ga 1$
based on a simple discontinuous model for QSO activity,
in which there is two short gas consumption events
during the first central inflow episode.
We scaled the QSO LF to the present day elliptical LF,
assuming that all \el s have harboured a QSO
during their \ev.
Both the starburst and the supermassive black hole models
predict the right shape of the QSO LF,
but the nuclear starburst
systematically underestimates the density  number of QSOs.
In addition, our model reproduces the \ev\ of the LF between
$z=1.25$ and $z=2.9$.
In our models, the mass deposited by the first central inflow
represents 0.15-0.5 \% of the present day luminous mass of the \gal.
Assuming that this mass goes to the formation of a central object
(star cluster or black hole), the model correctly predicts
for the Dark Massive Objects (DMOs) in the nuclei of  \gals,
both the masses and the DMO-to-galaxy mass ratios.

Another conclusion derived from our models is that the hosts of high-redshift
AGN should be relatively mature objects.
The calculated \ev\ of the inner \cf\ and energetic considerations
imply that the gas deposited by the inflow in the nucleus
should accumulate during $\sim 1$ Gyr 
before triggering a short-lived AGN activity event.
On the other hand, except for extremely large \gals\ (present-day $M_B=-24$),
the first, massive \cf, responsible for maintaining the AGN activity,
lasts for $2-3$ Gyr. 
In this scenario, if the high redshift \gal\ is to display strong AGN
activity, its probable age would range from $\sim 1$ to $\sim 3$ Gyrs.
The minimum age of $\sim 1$ Gyr for QSO hosts derived above is
consistent with the  $\sim 1$ Gyr time scale for metal enrichment,
needed to explain the strong metal lines observed even in $z\sim 5$ QSOs.

One further prediction of our models is that,
at the present, only the most massive objects should be host to powerful AGN,
but at high redshift powerful AGN activity is expected even  
for hosts of smaller mass.
Interestingly enough, this is what seems to be observed,
for radio \gals\ at least.
Models 1 and 2 are examples of sub-$L^*$ \gals\ with strong AGN activity
at high redshift.
Note that $\zeta$ increases with decreasing \gal\ mass for masses below
$M_G \approx 10^{12}$ \msun.
Since this parameter describes the relative importance of the first \cf\ episode
occuring when the \gal\ is less than $2-3$ old, 
this means that for the lower mass systems,
the efficiency of building-up the nucleus is higher than in larger systems,
and that they have an early \cf\ massive enough
to sustain a strong AGN activity.
However, this activity is limited only to the 2-3 first Gyr of the \gal\ and,
therefore, smaller systems with strong nuclear activity are to be
found only at high redshift.
Even in the case of the episodic scenario of model 2(1/3), 
the late nuclear activity is much weaker at low redshift 
(in this model, the two late central inflow episodes deposit
only 1.8 \% and 1 \% of the mass of the first inflow episode).
On the other hand, models 10, 20 and 50 exhibit
a present-day massive central \cf\ which could trigger AGN activity.
In these massive systems, the late cooling flow may accumulate
into the nucleus an amount of mass comparable 
to that of  the early \cf\  (model 20 is 
typical of this case, the late \cf\ deposits
$1.65\times 10^9$ \msun, i.e. 47 \% of the mass of the first \cf).
Only these objects, therefore, would harbour, intense AGN activity
at low redshift.
Support to this picture is given by recent analysis of host \gals\ of powerful
nearby ($z \la 0.3$) AGN, belonging to three samples --- radio \gals\, 
radio-loud quasars, and radio-quiet  quasars (Taylor et al. 1996).
For all three classes of AGN, the host \gals\ are 
large (half-light radius $r_{1/2} \geq 10$ kpc)
and luminous (K-band luminosity $L_K\geq L^*$).
(Note that the less massive model exhibiting a present-day central is model 10, 
with $M_B=-21.8$, or $L_B=2 L^*$.)

Finally, we should note that, although pure luminosity \ev\ seems to reproduce
the \ev\ of the QSO LF, in our models the individual QSOs do not
dim over cosmological time scales, but rather are
short-lived (i.e., $t_{on}=1-3\times 10^7$ yr).
In order to comply with the energetic requirements of the QSOs,
their activity must occur in short episodes of massive 
consumption of mass accumulated in the galactic
nucleus during a much longer span of time ($\sim 1$ Gyr).
This sort of episodic activity displayed by our model for QSOs
seems to be a rule among the AGN in general,
since for other class of AGN, the radio galaxies, the radio LF also seems
to follow pure luminosity \ev, and yet the 
radio sources themselves seem to have lifetimes of only a few $10^7$ yr.
As a matter of fact, a comparison between the properties of the radio-loud
population and the present model would be very useful
to clarify the relation of the QSO phenomenon and the early \ev\ of
\el\ \gals, since radio observations allow us to explore
the $z>2$ domain without the need of uncertain corrections
for dust absorption and lensing bias
that hamper the optical tecniques.

\section*{Acknowledgments}

A.C.S.F. acknowledges support from the Brazilian agency CNPq.
We thank the referee James Dunlop for a number of suggestions which helped
us to improve this paper significantly and specially for the remarks on
the radio loud population discussed in the two last paragraphs of this paper.

\bsp

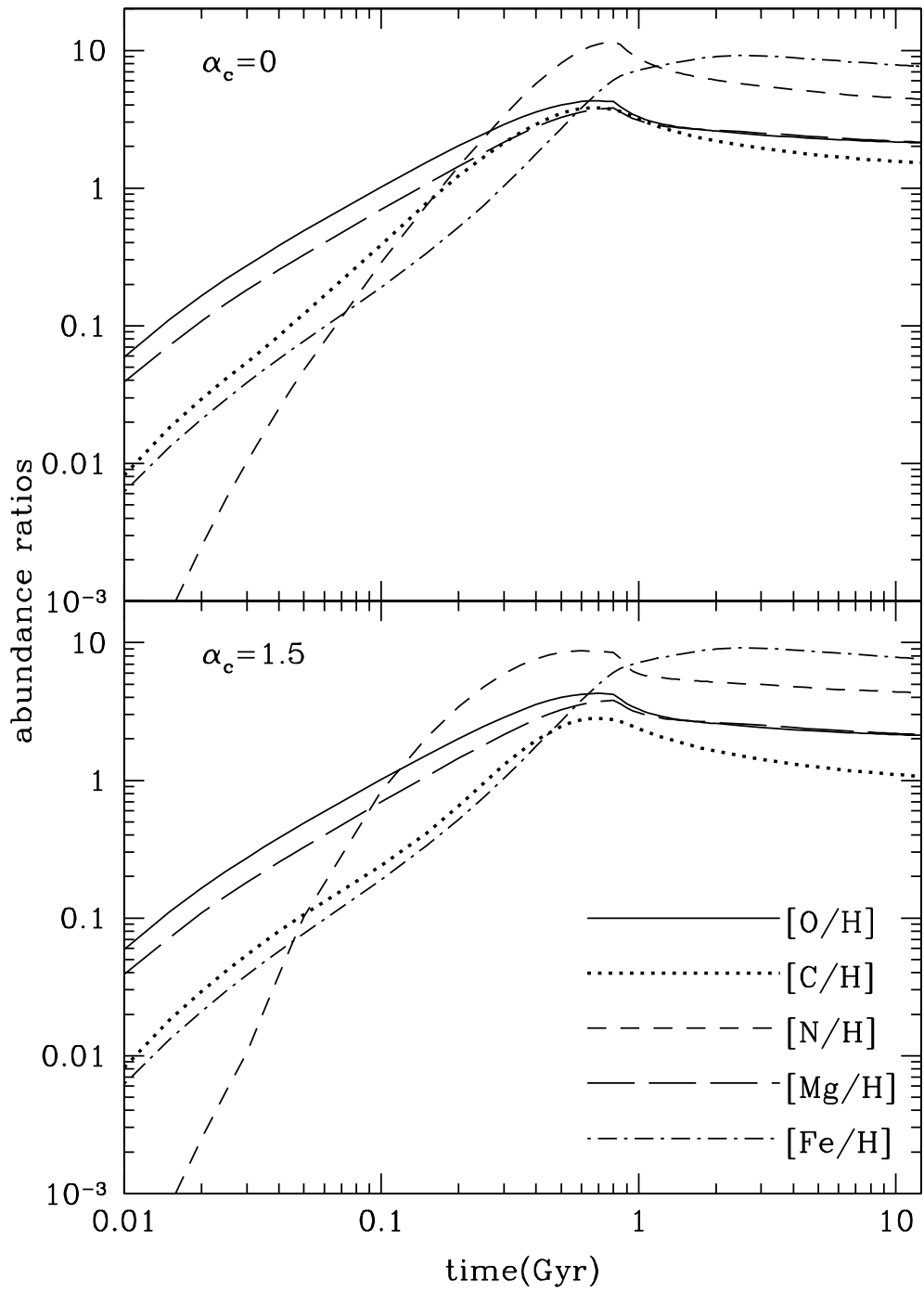
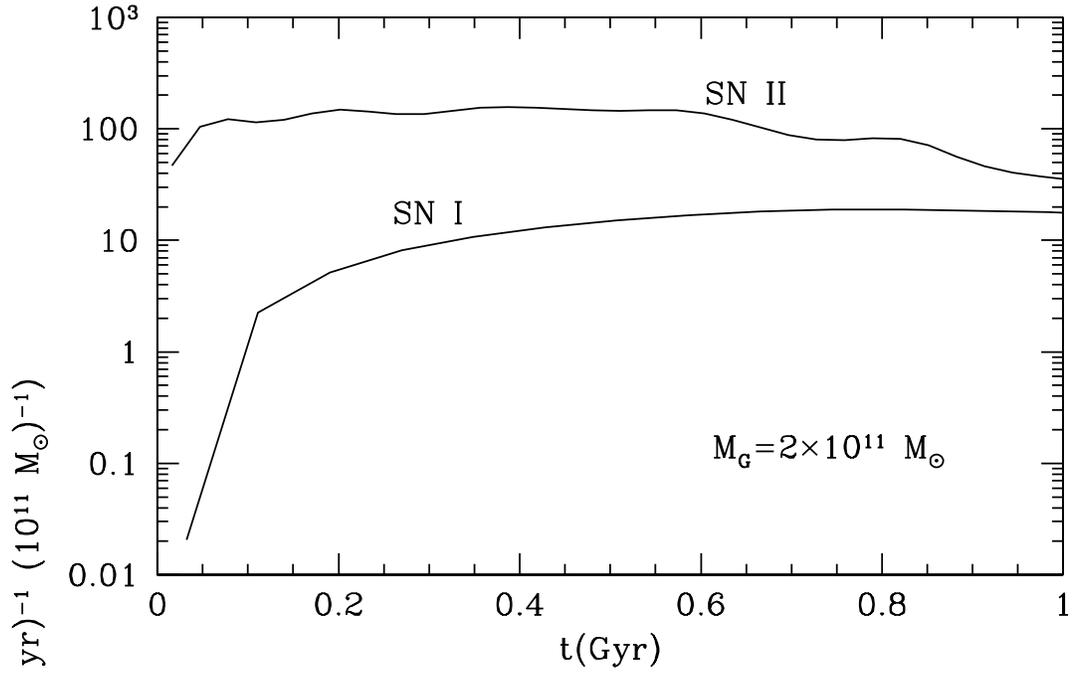
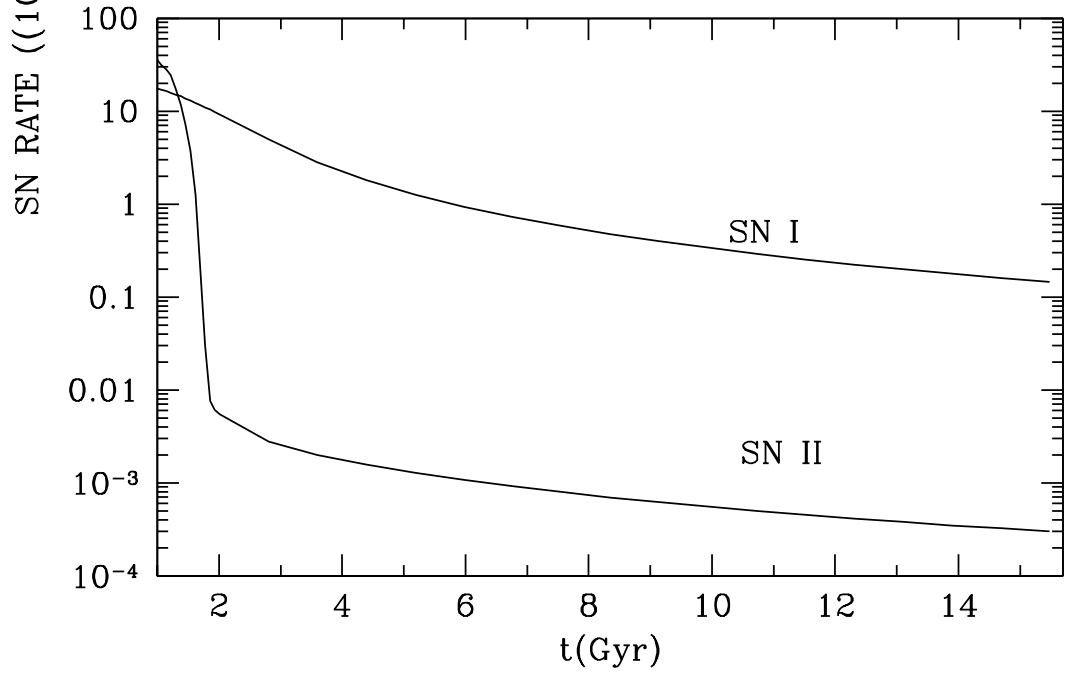
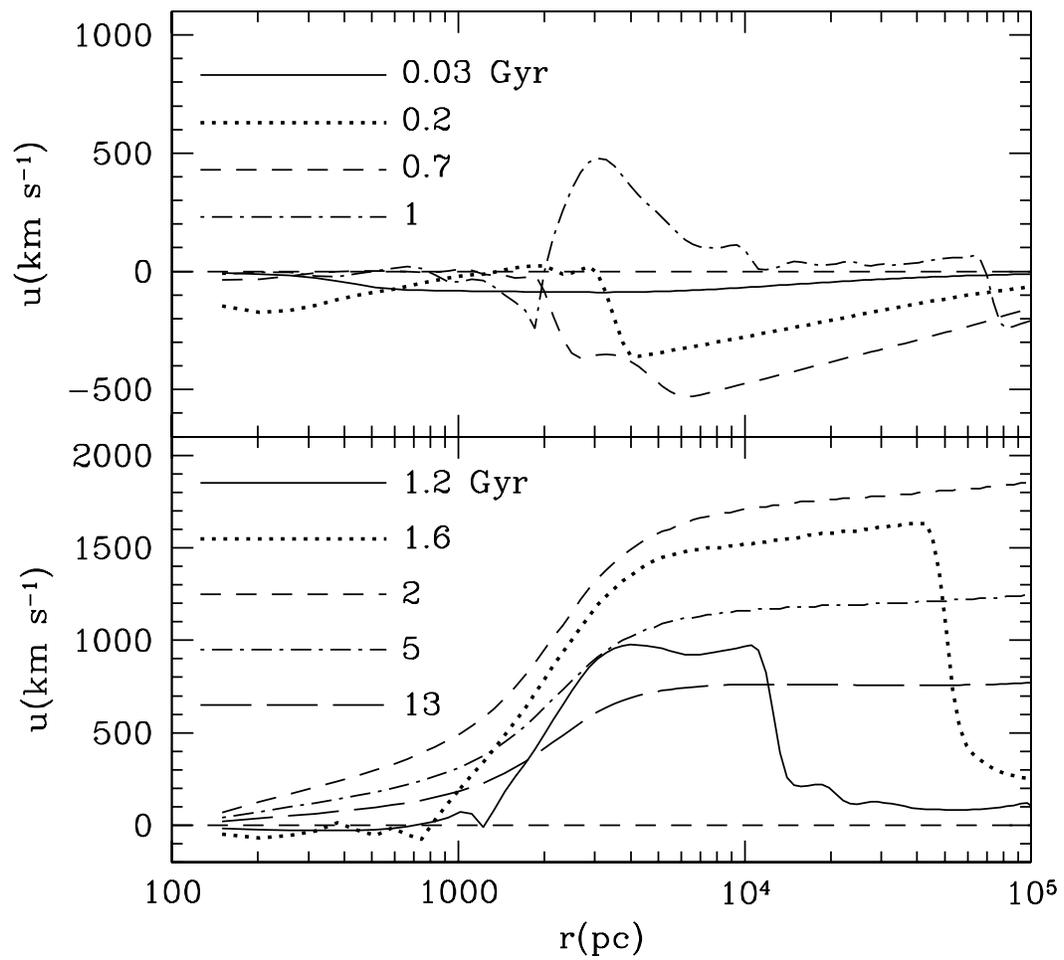
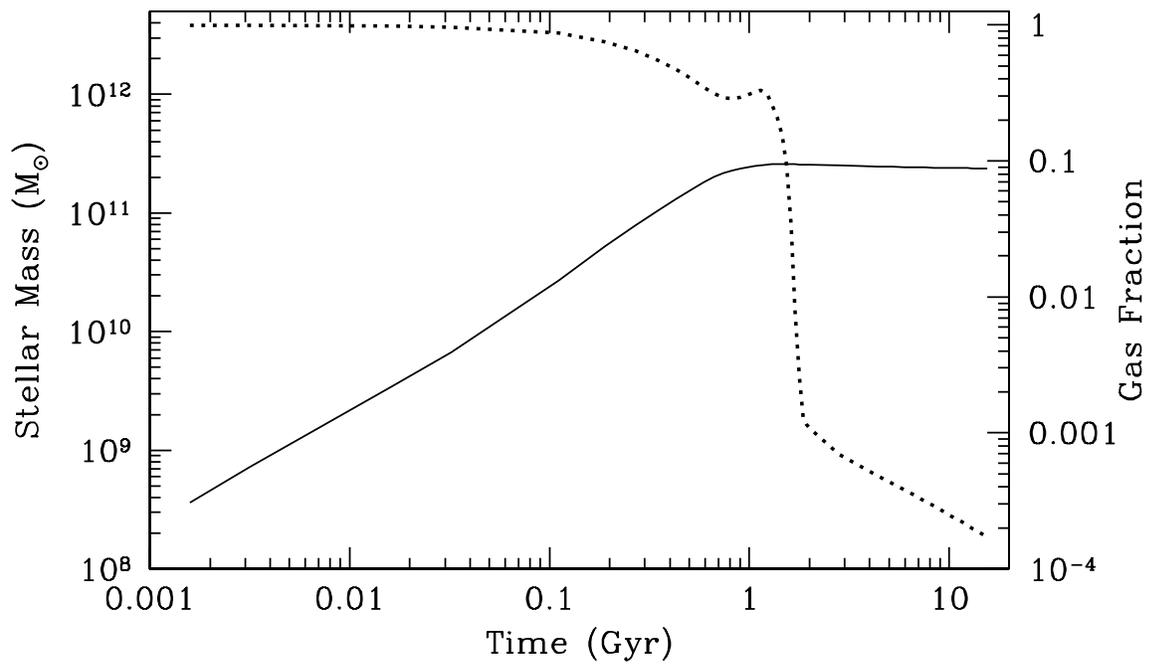
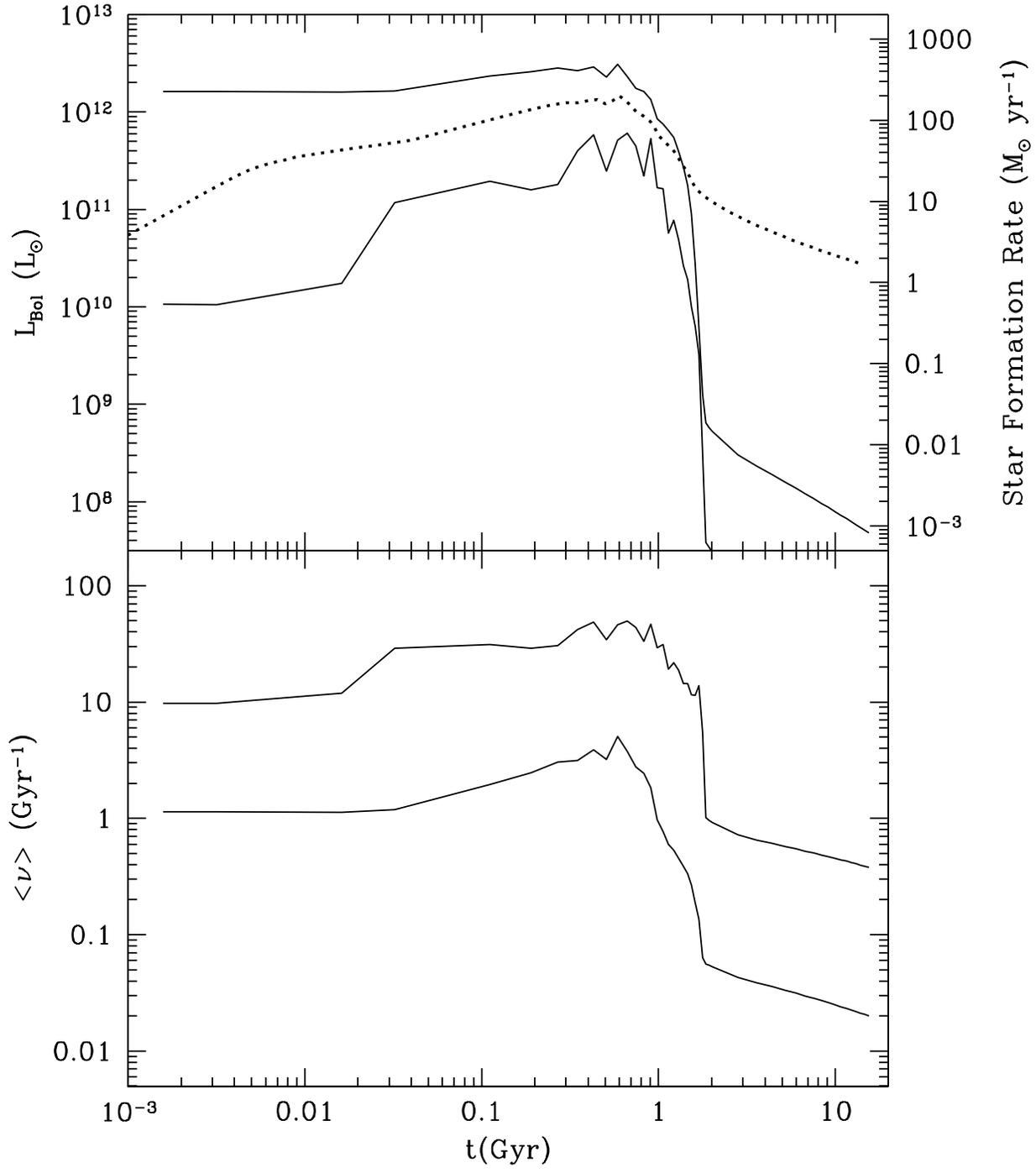
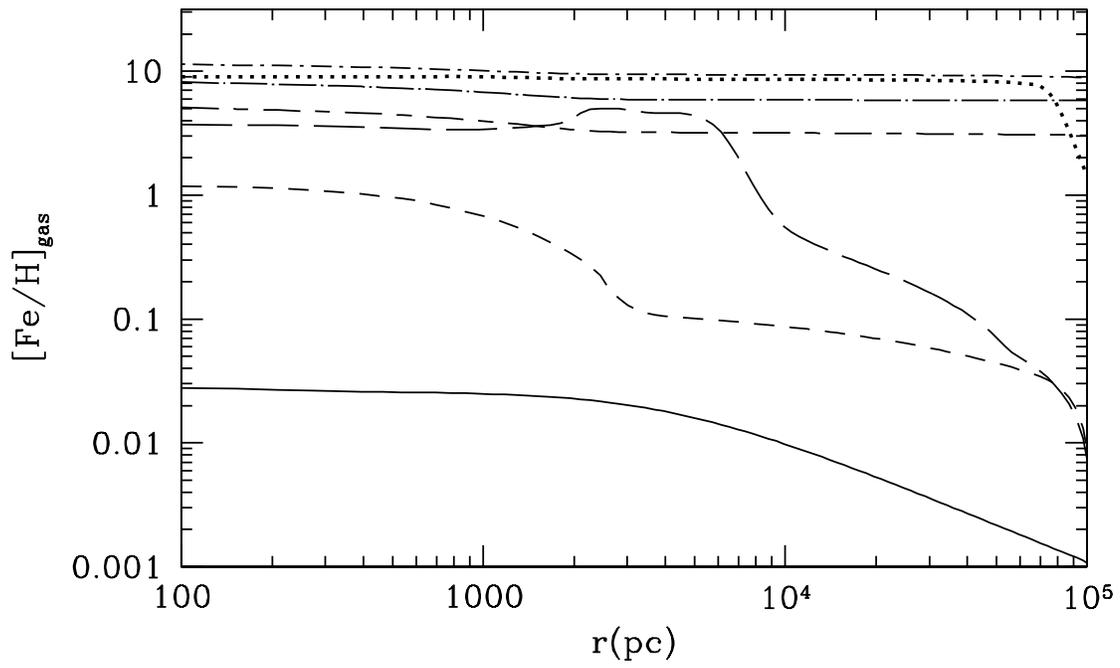
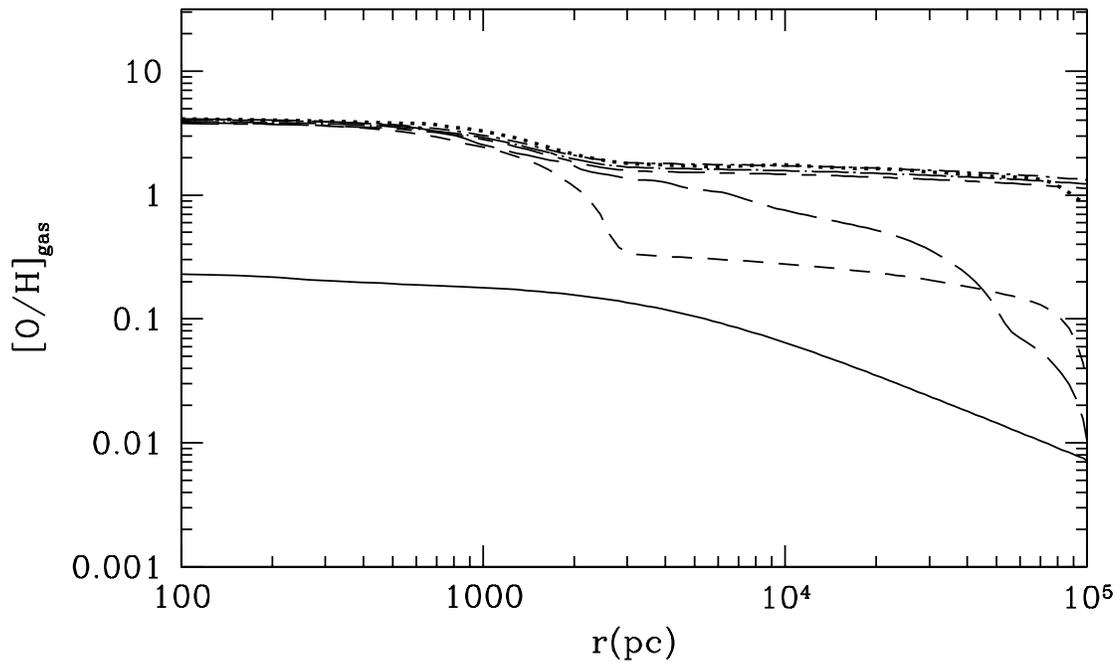
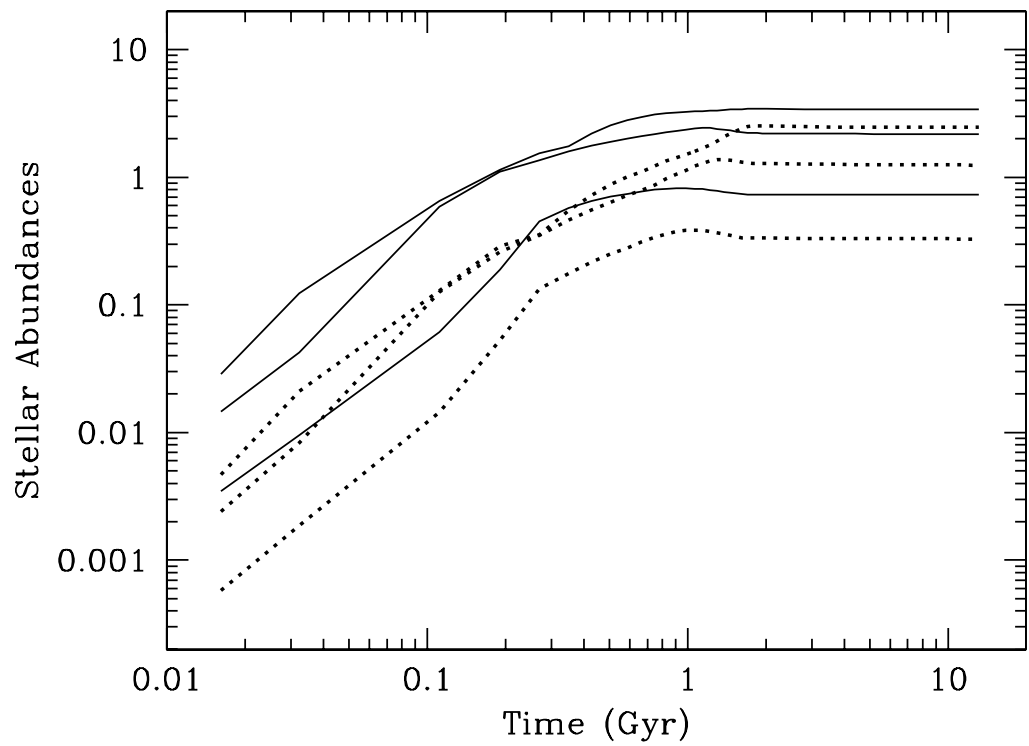
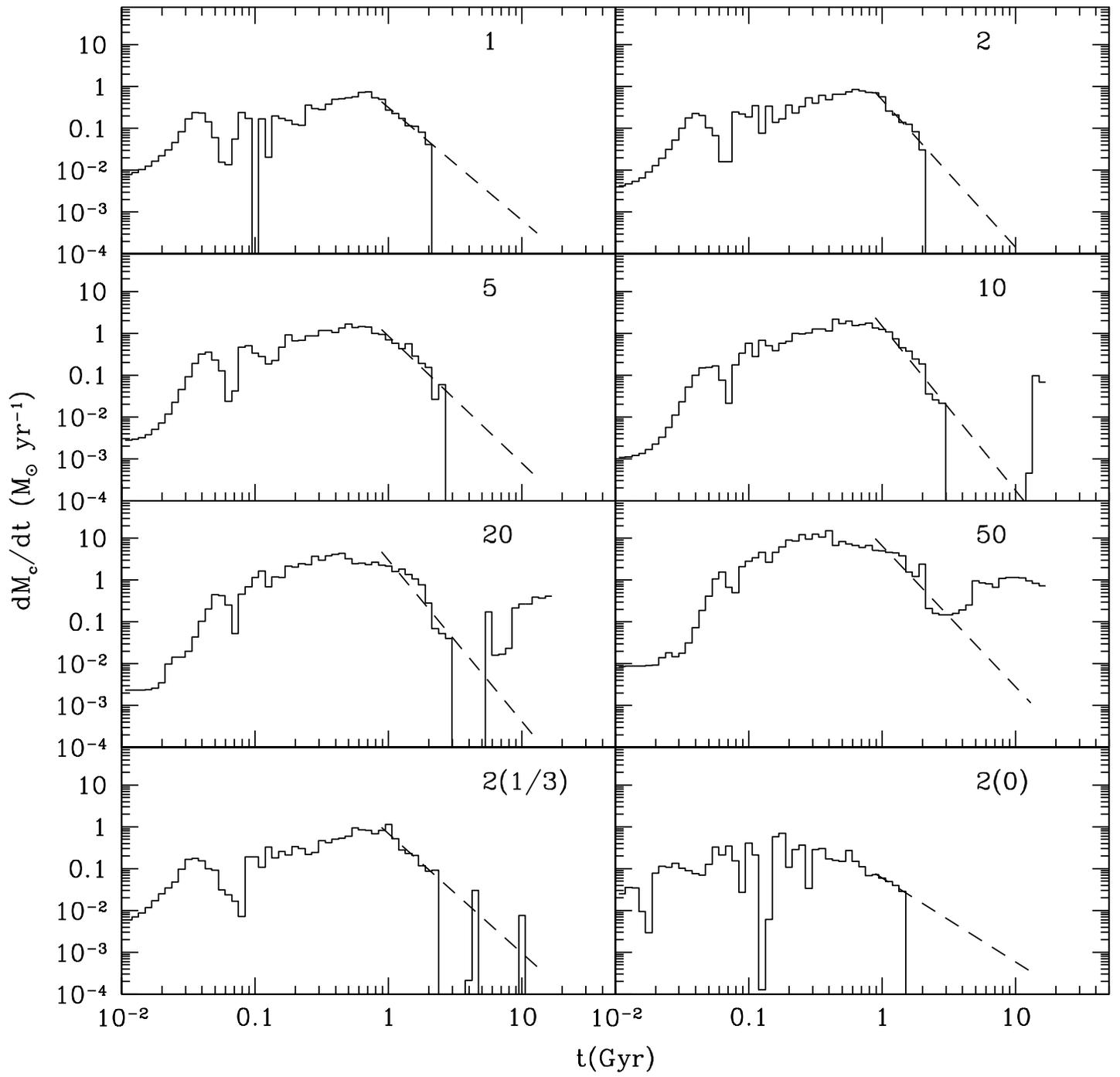
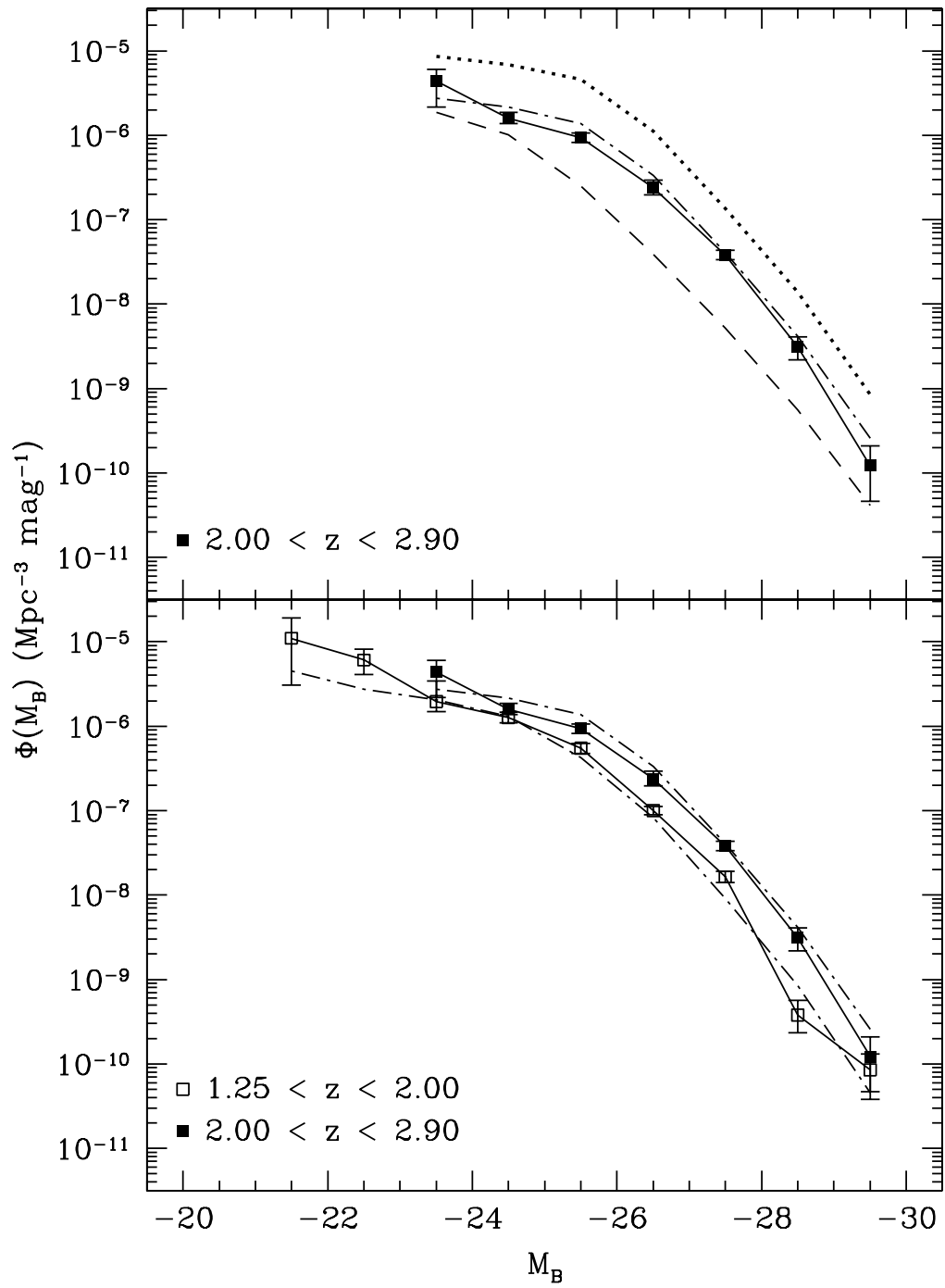
\end{document}